\documentclass[aps,prb,twocolumn]{revtex4-2}
\usepackage{graphics}
\usepackage{booktabs}
\usepackage{caption}
\usepackage{subcaption}
\usepackage{epstopdf}
\usepackage{epsfig}
\usepackage[flushleft]{threeparttable}
\usepackage{placeins}
\usepackage{epsf,epic}
\usepackage{color}
\usepackage{marvosym }
\usepackage{amsmath}
  \DeclareMathOperator{\Tr}{Tr}
\usepackage[utf8]{inputenc}
\usepackage{pgfplots}
\pgfplotsset{compat=newest}
\usepackage{amssymb}
\usepackage{amsfonts}
\usepackage{wrapfig}
\usepackage{multirow}
\usepackage{bm}
\usepackage{dcolumn}
\newcommand{\etal}{\textit{et al.\ }}
\newcommand{\ie}{\textit{i.e.\ }}
\newcommand{\eg}{\textit{e.g.\ }}
\graphicspath{{Figs/}}
\begin{document}
\title{Real space representation of the quasiparticle self-consistent
  $GW$ self-energy and its application to
defect calculations.}
\author{Ozan Dernek}
\author{Dmitry Skachkov}
\altaffiliation{Current address: Department of Physics, University of Florida, Gainesville, Florida 32611, USA}
\author{Walter R. L. Lambrecht}
\affiliation{Department of Physics, Case Western Reserve University, 10900 Euclid Avenue, Cleveland, OH 44106-7079, USA}
\author{Mark van Schilfgaarde}
\affiliation{National Renewable Energy Laboratory, Golden, CO 80401, USA}

\begin{abstract}
  The quasiparticle self-consistent QS$GW$ approach incorporates the corrections of the
  quasiparticle energies from their Kohn-Sham density functional theory (DFT) eigenvalues
  by means of an energy independent and Hermitian self-energy matrix usually given in the
  basis set of the DFT eigenstates. By expanding these into an atom-centered basis set
  (specifically here the linearized muffin-tin orbitals) a real space representation of
  the self-energy corrections becomes possible. We show that this  representation is
  relatively short-ranged. This offers new opportunities to construct the self-energy of
  a complex system from parts of the system by a cut-and-paste method. Specifically for a
  point defect, represented in a large supercell, the self-eneregy can be constructed from
  those of the host and a smaller defect containing cell. The self-energy of the
  periodic host can be constructed simply from a $GW$ calculation for the primitive cell.
  We show for the case of the As$_\mathrm{Ga}$ in GaAs that the defect part can already
  be well represented by a minimal 8 atom cell and allows us to construct the self-energy
  for a 64 cell in good agreement with direct QS$GW$ calculations for the large cell.
  Using this approach to an even larger 216 atom cell shows the defect band
  approaches an isolated defect level. The calculations also allow to identify a second
  defect band which appears as a resonance near the conduction band minimum. The results
  on the extracted defect levels agree well with Green's function calculations for an
  isolated defect and with experimental data. 
\end{abstract}
\maketitle

\section{Introduction}
  In recent years, many-body-perturbation theory, specifically in Hedin's $GW$
  approximation\cite{Hedin65,Hedin69}, where $G$ is the one-electron Green's function and
  $W$ the screened Coulomb interaction, has emerged as the method of choice to calculate
  meaningful quasiparticle excitation energies as opposed to Kohn-Sham density functional
  one-electron energies. In particular, this yields more accurate band structures and band
  gaps in good agreement with experiment, typically within $\sim$0.1 eV depending
  somewhat on the details of the implementation and the material.\cite{MvSQSGWprl}

  For localized levels, such as point defects, on the other hand, excitations are usually
  calculated by means of a $\Delta$SCF approach from differences between two total
  energies. More precisely for point defects, it is now standard procedure to calculate
  the energies of formation as function of Fermi level position (\ie the electron chemical
  potential) in the gap and then to determine the transition energies, which are the
  crossing points of these energies of formation from one charge state to the other. On
  the other hand, it has been pointed out that the $GW$ quasiparticle energies for a
  defect system can be directly related to ``vertical'' excitations, meaning excitations
  which keep the structure unchanged. Apart from excitonic effects, their differences
  represent the optical transitions for example for transferring an electron from a defect
  level to the conduction band or from the valence band to a defect
  level.\cite{Rinke09,Lany10GWVO} For example, transferring an electron from a defect
  level to the conduction band minimum (CBM), changes the defect from one charge state
  $q$ to another $q+1$, with an electron in a delocalized conduction band state. Thus the
  difference in CBM and defect level quasiparticle energies calculated at the fixed
  geometry of the $q$ state, is the ``vertical'' excitation energy for this process.
  Subsequently, one may add relaxation energies of the defect to lowest energy geometry
  within a given charge state. Thus, the thermodynamic transition state level of the
  $q/q+1$ transition may be obtained from a total energy relaxation within a given charge
  state calculated at the DFT level combined with the quasiparticle excitation energies
  calculated at the $GW$ level. Apart from the excitonic effect, the ``vertical''
  transition is often directly of interest as an approximation to the optical transition.
  In practice, defect levels are usually calculated in supercells using periodic boundary
  conditions, and the defect levels turn into defect bands. So, the above relation gives
  a renewed incentive to take defect one-electron band structures seriously, rather than
  dismissing them as irrelevant Kohn-Sham eigenvalues, provided they are calculated at the
  $GW$ quasiparticle level.

  Nonetheless, the $GW$ method has not yet found widespread applications in defect
  studies. This is at least in part due to the large computational effort required for
  $GW$ calculations. In particular, the latter is challenging for the large supercells
  needed to represent defects adequately. Thus, there is a need for improving the
  efficiency of the $GW$ approach, eventually at the cost of some simplification, to make
  it applicable to larger systems.

  On the other hand, in many defect calculations, the infamous gap underestimate of the
  typical semi-local generalized gradient approximation (GGA) or local density
  approximations (LDA), can lead to serious errors in defect calculations. Defect levels,
  which should be in the gap may end up in the band continuum. This also affects the total
  energies and therefore transition levels if one considers charge states in which that
  defect level (now a resonance in the band) is given an extra charge because in the
  calculation that charge is actually placed in a delocalized state at the bottom of the
  band rather than in the defect level. Thus there seem to be some advantages to start
  from a more accurate quasiparticle one-electron theory such as $GW$. Total energies with
  in the QS$GW$ approach are in principle calculable at the random phase approximation
  (RPA) level by means of an adiabatic connection approach\cite{Kotani07}, but are
  difficult to converge because they require a sum over unoccupied bands.

  It would seem desirable to at least build into the calculation of the defect, the
  correct band structure of the host, overcoming thereby the band gap problem. Several
  such approaches have been used in the past: for example, Christensen\cite{Christensen84}
  advocated using a $\delta$-function corrected potential in which a $\delta$-function
  placed at the cation site and some interstitial sites, raises the $s$-like partial waves
  at those sites in energy and since these $s$-like orbitals form a predominant part of
  the conduction band, they artificially corrected the gap. LDA+U with several orbital
  dependent $U$ parameters were also shown to provide an effective way to mimic the
  corrected band structure.\cite{Paudel08VO,Boonchun2011,Skachkov16} Non-local external potentials
  (NLEP) of the form $\Delta V_{\alpha,l}^\mathrm{NLEP}$ adjusted to reproduce the
  conduction band structure were introduced by Lany and Zunger.\cite{LanyNLEP} Modified
  pseudopotential were also used for this purpose.\cite{Segev07}

  The question thus arises to what extent we can decouple the host band structure effects
  from the defect. Our goal with this project was to explore whether the $GW$ self-energy
  of a defect system could be constructed from that of the host and the defect site itself
  or its immediate neighborhood without having to carry out the expensive full $GW$
  calculation for the large unit cell required to adequately represent a defect.

  The most prevalent approach nowadays to incorporate the gap corrections beyond
  semi-local functions, is to use a hybrid functional such as the Heyd-Scuseria-Ernzerhof
  (HSE).\cite{HSE03,HSE06,PBEh} That approach significantly improves the gaps by including
  a fraction of the exact exchange operator cutoff usually beyond a certain range. By
  adjusting the fraction of the exchange included, the gap can be adjusted. While this
  approach correctly incorporates the gap correction, it is less clear that it also
  adjusts both band edges individually and/or obeys the generalized Koopmans theorem (GKT)
  \cite{Lanypolaron,LanyGKTNO} for different defects simultaneously with the host band
  structure. Furthermore it is also a relatively expensive approach with  computational
  effort well beyond that of a semi-local calculation. The approach we present here to
  construct the $GW$ self-energy is also applicable to the non-local exact exchange and
  hence, could also make that approach more efficient.

  In this paper, we show that the QS$GW$ self-energy matrix can be expanded in
  atom-centered orbitals, such as the linearized muffin-tin orbitals (LMTO). If these are
  chosen sufficiently localized, then the self-energy matrix can be represented in real
  space within a finite range. One might envision doing this also with maximally localized
  Wannier functions. This then offers new opportunities to approximately construct the
  self-energy of a system by partitioning the system in sub parts and constructing the
  self-energy by a cut-and-paste approach. In particular, we apply this here to point
  defects. We first construct the self-energy of the host in a supercell from that of the
  primitive cell. In a second step, we replace the part of the self-energy matrix related
  to the defect atom and its near neighborhood in terms of the self-energy of a smaller
  supercell containing the defect for which a GW calculation is more readily feasible. We
  validate the accuracy of the approach with the well-studied
  case of the As$_\mathrm{Ga}$ defect.

\section{Computational Approach}
\subsection{GW background}
  In many-body-perturbation theory quasiparticle excitation energies are given by the
  equation
    \begin{eqnarray}
      \left[  -\frac{1}{2}\nabla^2 + v_N({\bf r})+ v_H({\bf r})\right]\Psi_i({\bf r})
      \nonumber \\
      +\int d^3r' \Sigma_{xc}({\bf r},{\bf r}',E_i)\Psi_i({\bf r}')=E_i\Psi_i({\bf r}),
      \label{eqqp}
    \end{eqnarray}
  where we use Hartree atomic units ($\hbar=e=m_e=1)$, $v_N$ is the nuclear potential and
  $v_H$ the  Hartree potential, $\Sigma_{xc}$ the exchange-correlation self-energy and $\Psi_i$ is the quasiparticle wave function. In
  the $GW$ approximation, the latter is calculated from $\Sigma(12)=iG(12)W(1^+2)$ where
  $1$ is a shorthand for $\{{\bf r}_1,\sigma_1,t_1\}$, \ie position, spin and time of the
  particle 1, $1^+$ means $\lim_{\delta\rightarrow+0} t_1+\delta$,  $G(12)$ is the
  one-electron Green's function and $W(12)$ the screened Coulomb interaction. The exact
  one particle Green's function is defined by
  $G(12)=-i\langle N|T[\psi(1),\psi^\dagger(2)]|N\rangle$, with $T$ the time-ordering
  operator, $\psi(1)$ the annihilation field operator and $|N\rangle$ the $N$-electron
  ground state. The screened Coulomb interaction is given by
  $W(12)=v(12)+\int d(34) v(3)P(34) W(42)$ and $P(12)=-iG(12)G(21)$ is the irreducible
  polarization propagator. In practice, it is usually obtained starting from some
  effective independent-particle Hamiltonian,
    \begin{equation}
      H^0=  -\frac{1}{2}\nabla^2 + v_N({\bf r})+ v_H({\bf r})+v_{xc}({\bf r})
    \end{equation}
  with for example the local density approximation (LDA) exchange-correlation potential
  $v_{xc}({\bf r})$. The Green's function $G^0$ is then constructed from the eigenvalues $\epsilon_i$ and
  and eigenfunctions $\psi_i$ of
    \begin{equation}
      H^0\psi_i({\bf r})=\epsilon_i\psi_i({\bf r})
    \end{equation}
  as follows
    \begin{equation}
      G^0({\bf r},{\bf r}',\omega)=\sum_i \frac{\psi_i({\bf r})\psi_i^*({\bf r}')}
      {\omega-\epsilon_i+i\delta sgn(\epsilon_i-\mu)},
    \end{equation}
    with $\mu$ the chemical potential and $\omega$ the energy variable of the
    Green's function.
  These LDA eigenstates form a convenient basis set in which the Green's function
  $G^0_{ij}(\omega)=\delta_{ij}(\omega-\epsilon_i\pm i\delta)^{-1}$ is diagonal. For a
  solid, the states are labeled by $i=\{n,{\bf k}\}$ with $n$ a band index and ${\bf k}$
  the point in the Brillouin zone.  The screened ($W^0$) and bare ($v$) Coulomb
  interactions and the polarization $P^0$ on the other hand are expressed in an auxiliary
  basis set of Bloch functions. In the LMTO implementation of the $GW$ method, these are
  constructed from products of angular momentum partial waves\cite{Aryasetiawan94} inside
  the spheres and plane-waves confined to the interstitial space, which is afterwards
  reduced to avoid linear dependence and rotated so as to diagonalize the bare Coulomb
  interaction.\cite{Kotani07,Kotanijpsj,Friedrich10} We label them
  $E_{{\bf q}\mu}({\bf r})$ and the matrix of the Coulomb interactions in terms of them is
  then written
    \begin{equation}
      v_{\mu\nu}({\bf q})=\int d^3r d^3r' E_{{\bf q}\mu}^*({\bf r})
      \frac{1}{|{\bf r}-{\bf r}'|}E_{{\bf q}\nu}({\bf r}')
    \end{equation}
  and
    \begin{equation}
      W^0_{\mu\nu}({\bf q},\omega)=\left[ 1- v_{\mu\lambda}({\bf q})
        P^0_{\lambda\kappa}({\bf q},\omega)\right]^{-1}v_{\kappa\nu}({\bf q})
    \end{equation}
  is obtained from a matrix inversion once $P^0_{\mu\nu}({\bf q},\omega)$ is known. The
  latter is also calculated directly in terms of the $\psi_{n{\bf k}}$ and eigenvalues
  $\epsilon_{n{\bf k}}$.\cite{Kotani07} In the above equation, summation convention over
  repeated indices is understood. In most other works, plane waves are used instead as
  basis functions. Let's write the rotation from the auxiliary functions $E^{\bf q}_\mu$
  to the LDA eigenstates as
  $\langle \psi_{{\bf k}n}|\psi_{{\bf k}-{\bf q}n'}E^{\bf q}_\mu\rangle$. Using these
  rotation matrix we can express $W$ as:
    \begin{eqnarray}
      W^0_{nn'm{\bf k}}({\bf q},\omega)=\sum_{\mu\nu}
      \langle\psi_{{\bf k}n}|\psi_{{\bf k}-{\bf q},n'}E^{\bf q}_\mu\rangle
      W^0_{\mu\nu}({\bf q},\omega)\nonumber \\
      \langle E_\nu^{\bf q}\psi_{{\bf k}-{\bf q}n'}|\psi_{{\bf k}m}\rangle
    \end{eqnarray}
  The self-energy matrix is then given by
    \begin{eqnarray}
      \Sigma_{nm}({\bf k},\omega)=\frac{i}{2\pi}\int d\omega' \sum_{\bf q}\sum_{n'}
      G^0_{nn'}({\bf k}-{\bf q},\omega-\omega') \nonumber \\
      W^0_{nn'm;{\bf k}}({\bf q},\omega')e^{i\delta\omega'}
    \end{eqnarray}
  in which we recognize the schematic $\Sigma=iGW$ but which makes it clear that to
  obtain the energy and {\bf k}-space dependent form, a triple convolution is involved over energy $\omega'$, ${\bf q}$ and band index $n'$.

  This self-energy matrix is energy dependent and contains in principle information, not
  only on the quasiparticle energies but also the satellites and non-coherent parts of
  the one-electron excitation. In the QS$GW$ method, we now reduce this to a non-local but
  energy-independent and Hermitian matrix,
    \begin{equation}
      \tilde{\Sigma}_{nm}({\bf k})=\frac{1}{2}\mathrm{Re}
      \left[\Sigma_{nm}({\bf k},\epsilon_{n{\bf k}})
        +\Sigma_{nm}({\bf k},\epsilon_{m{\bf k}})\right]
    \end{equation}
  We now define
  $\Delta\tilde{\Sigma}_{nm}({\bf k})=\tilde{\Sigma}_{nm}({\bf k})-v^{xc}_{nm}({\bf k})$
  where we subtract the matrix element of the LDA  exchange-correlation potential taken
  between the Bloch eigenstates. We can then add this correction to the
  exchange-correlation potential to the $H^0$ Hamiltonian and re-diagonalize the latter to
  find new independent particle eigenvalues and eigenstates and repeat the cycle of
  calculating $\tilde{\Sigma}$. At the convergence of this iteration, the eigenvalues of
  the final  $H^0$, which we will call $H^\mathrm{QSGW}$ are identical to the
  quasiparticle energies. We may view this as finding  a $G^0W^0$ perturbation theory
  solution of Eq.(\ref{eqqp}) starting from $H^0$ but refining $H^0$ so the perturbation
  becomes negligible. In this sense the quasiparticle energies are self-consistent and
  independent of the starting approximation but they are still real and do not provide
  information on the lifetime of the actual quasiparticle states.

\subsection{Bloch-function and real space representation}
  The Bloch functions of the $H^\mathrm{QSGW}$ are now known as an expansion in the LMTO
  basis set.
    \begin{equation}
      |\psi_{n{\bf k}}\rangle=\sum_{{\bf R}i} |\chi_{{\bf R}\i}^{\bf k}\rangle
        b_{n{\bf R}i}^{\bf k}
    \end{equation}
  so we can re-express the self-energy correction matrix as
  $\Delta\tilde{\Sigma}_{{\bf R}i,{\bf R}'i'}({\bf k})$. Here {\bf R} label the sites in
  the unit cell and $i$ the muffin-tin orbitals. The latter are labeled by angular
  momentum  quantum numbers $(l,m)$ as well as a third index, labeling the choice of
  smoothed Hankel function decay and/or local orbital (confined to the muffin-tin sphere).
  See Ref. \onlinecite{Kotani07} for a full description of the full potential (FP)-LMTO method used.
  Finally, performing an inverse Bloch sum or Fourier transform, we obtain
  $\Delta\tilde\Sigma_{{\bf R},i;{\bf R}'+{\bf T},i'}$ fully expressed in real space,
  where ${\bf T}$ are the lattice translation vectors.

  Next, let us consider the same self-energy of the bulk system but represented in a
  supercell. Obviously there is a one-to-one mapping
  $\{{\bf R},{\bf T}\}\leftrightarrow\{{\bf R}_S,{\bf T}_S\}$ between positions ${\bf R}$
  in the primitive cell with lattice vectors ${\bf T}$ to the positions inside the
  supercell, ${\bf R}_S$, and the superlattice's lattice vectors, where
  ${\bf T}_S=\sum_i^3 n_i{\bf A}_i$ with $n_i$ integers, and the superlattice is defined
  by ${\bf A}_i=\sum_j N_{ij}{\bf a}_j$ with $N_{ij}$ a set of integers. Thus, we can
  obtain $\Delta\tilde\Sigma_{{\bf R}_S,i;{\bf R}_S'+{\bf T}_S,i'}$ by a simple relabeling
  procedure. In practice these are stored for every ${\bf R}_S$ using a neighbor table out
  to some maximum distance $|{\bf R}_S'+{\bf T}_S-{\bf R}_S|\le d_{max}$.

    \begin{figure}
      \includegraphics[width=8cm]{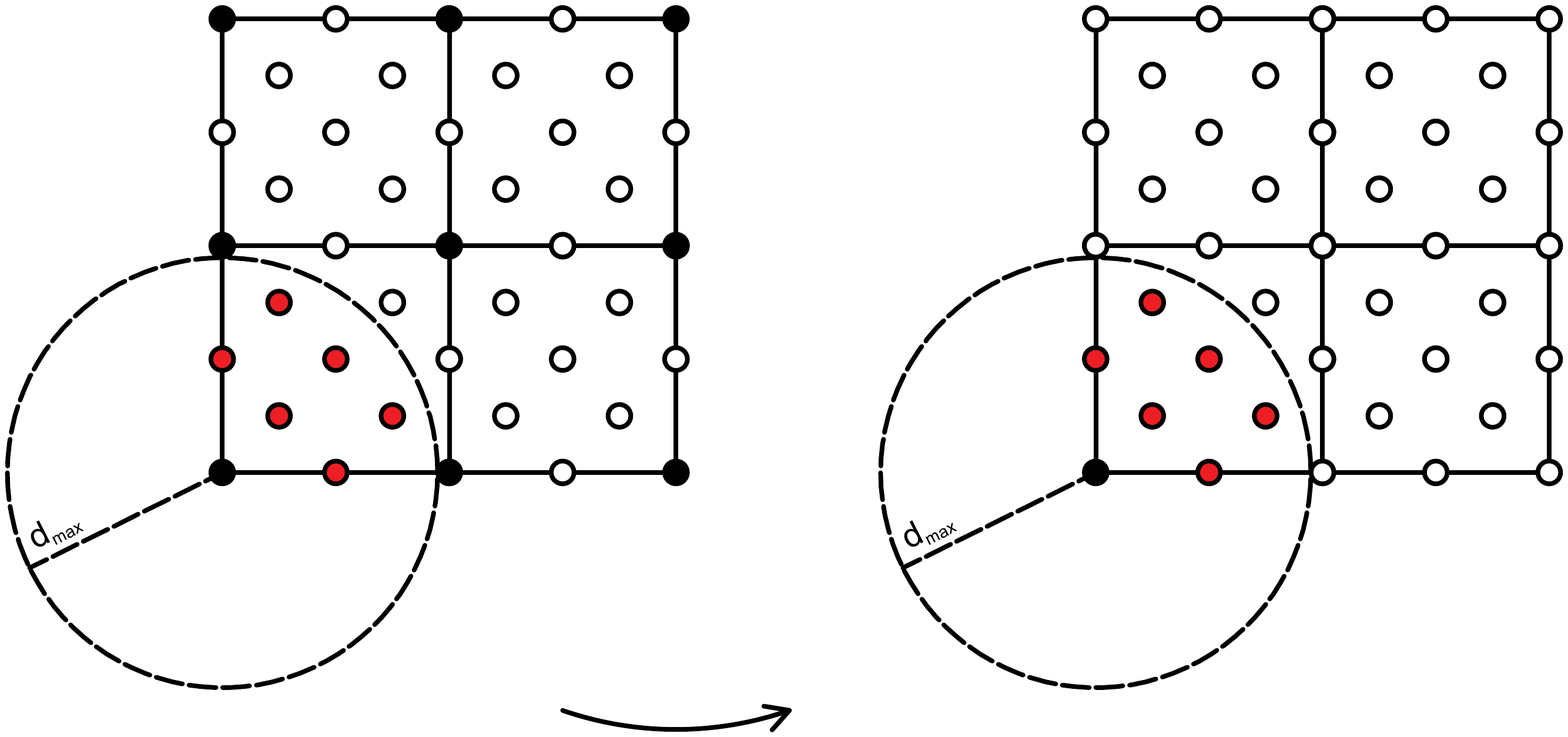}
      \caption{Schematic illustration of the self-energy editor cut-and-paste method. The
               left shows the $64^{8d}$ supercell with  defect atoms shown as black
               spheres and host atoms as open circles and the atoms within a range
               $d_{max}$ from the defect atom indicated by the dashed circle are red. The
               target system $64^{1d}$ with 1 defect is shown on the right. In this
               example, we assume only the defect atom itself comprises the defect region
               and the atoms within the range $d_{max}$ contribute to the self-energy in
               real space. The self-energy $\tilde\Sigma_{{\bf R},{\bf R}'}$ of the atom
               pairs corresponding to the red atoms connected to the defect atom in the
               target $64^{1d}$ cell are replaced by those from the $64^{8d}$ supercell,
               shown on the left.\label{figsupcel}}
     \end{figure}

\subsection{Self-energy cut-and-paste approach}
  The overall scheme for constructing the defect cell and its self-energy is as follows.
  For example, let us consider a 64 atom supercell to model the defect. We then start by
  creating the self-energy matrix of the perfect supercell (the host) from the self-energy
  in the primitive cell in the above form in real space and labeled according to the 64
  atom cell $\{{\bf R}_S,{\bf T}_S\}$ scheme. Once we have the self-energy matrix in the
  host supercell, we need to replace it by that for the defect within a certain range
  $d_{def}$ from the defect atom. For that purpose we construct a smaller supercell
  containing the defect, which we ultimately plan to use, \eg an 8 atom cell and carry out
  a self-consistent DFT calculation for it and subsequently a QS$GW$ calculation of its
  self-energy.  We then transform the self-energy of this defect cell again to a new 64
  atom cell by a similar relabeling step. Let us call this the $64^{8d}$ cell
  where 64 indicates the number of atoms in the cell and
  the superscript indicates
  the cell contains 8 defects. Next , we
  modify the host 64 atom cell by inserting the defect. We then carry out a
  self-consistent calculation for it at the DFT level and construct its self-energy by the
  following cut-and-paste method. The first step of the method is to create a blank place
  holder in memory to hold the self-energy around each atom, the type of atoms, and their
  orbitals at each of its neighbors. We then copy the corresponding self-energy into it
  from the host 64 atom cell and subsequently replace it by that of the defect $64^{8d}$
  cell for atoms within a certain distance $d_{def}$ from the defect site. The rest of the
  atoms are left unchanged as host atoms. Each copy step only happens according to the
  neighbor table up to a maximum range $d_{max}$. The copying of the self-energy orbital
  blocks happens by pairs. Only half of the $\Sigma_{{\bf R_S}i;{\bf R}_S'+{\bf T}_S,i'}$
  matrix elements need to be constructed because of the hermiticity. Once assembled in
  real space it can be Fourier transformed back according to the periodicity of the
  supercell, to find $\Sigma_{{\bf R}_Si;{\bf R}_S'i'}({\bf k}_S)$. Finally, we then need
  to carry out just one DFT self-consistent step in which the thus assembled  estimate of
  the self-energy is added to the $H^0$ DFT Hamiltonian and we can evaluate its band
  structure. The scheme is illustrated in Fig. \ref{figsupcel}.

  It is clear that for the scheme to work, $d_{max}$  must fit within the small defect
  containing supercell, so that the self-energy in the final cell for a pair of atoms of
  which one is within the defect range $d_{def}$ does not have a neighbor of the wrong
  type, in other words, it must still be a host like atom as in the final cell, not
  another defect atom.

  Furthermore, we need to allow for relaxation of the atoms near the defect. Therefore the
  atoms are mapped between the different cells based on their atom numbering and
  connectivity, not on the basis of their exact position. In principle, one could first
  relax the atoms in the large cell with the defect at the DFT level and then do the
  mapping to the perfect crystal and small defect cell even if their atomic positions do
  not perfectly match. Alternatively we could assume that the self-energy is not too
  sensitive to the relaxation and hence keep the self-energy fixed after our initial
  cut-and-paste operation and afterwards, relax (or relax again) the positions in the
  presence of that fixed self-energy. It is important here to remember that the
  self-energy provides only a correction to the electronic structure beyond the DFT
  Hamiltonian. The main defect induced changes are are already contained in
  at the DFT level.

\subsection{Computational details}
  The method has been implemented in the LMTO and QSGW suite of codes, named Questaal
  (Quasiparticle Electronic Structure and Augmented LMTOs)\cite{questaalpaper,questaal}.
  The basis sets are specified in terms of the angular momentum cutoffs and smoothed
  Hankel functions. For the initial tests on GaAs with the As$_\mathrm{Ga}$ antisite
  defect, we used a rather minimal basis set as specified along with the results. This
  leads to an overestimate of the band gap but is convenient for our present purpose of
  demonstrating the validity of the approach. Lattice positions are optimized for the
  crystal cells with defects. Details of the supercells chosen in the cut-and-paste
  approach are given along with the test results. In the final Sec. \ref{largebs} we use a larger basis set to achieve accurate comparison to experiment.

\section{Results}
\subsection{Gap convergence with self-energy real space cutoff} \label{gapcon}
  We start by testing the core idea of a finite range self-energy for bulk GaAs. Table
  \ref{tabgap} shows the band gap of GaAs in QS$GW$ as function of the  cutoff $d_{max}$
  used in the real space representation of the self-energy. We can see that as soon as
  $d_{max}/a>1$ with $a$ the lattice constant, or a cluster of about 30 atoms is included,
  the gaps become reasonable, although the convergence is not uniform and it takes till a
  cluster of about 450 atoms or $d_{max}/a\approx 3$ to get absolute convergence. We note
  that in an $8$ atom supercell the distance between defects is only 1 cubic lattice
  constant and thus we have to restrict the $d_{max}/a\approx 1$, but nonetheless, this
  already will be shown to give quite reasonable results for the self-energy.

\begin{figure}
      \includegraphics[width=8cm]{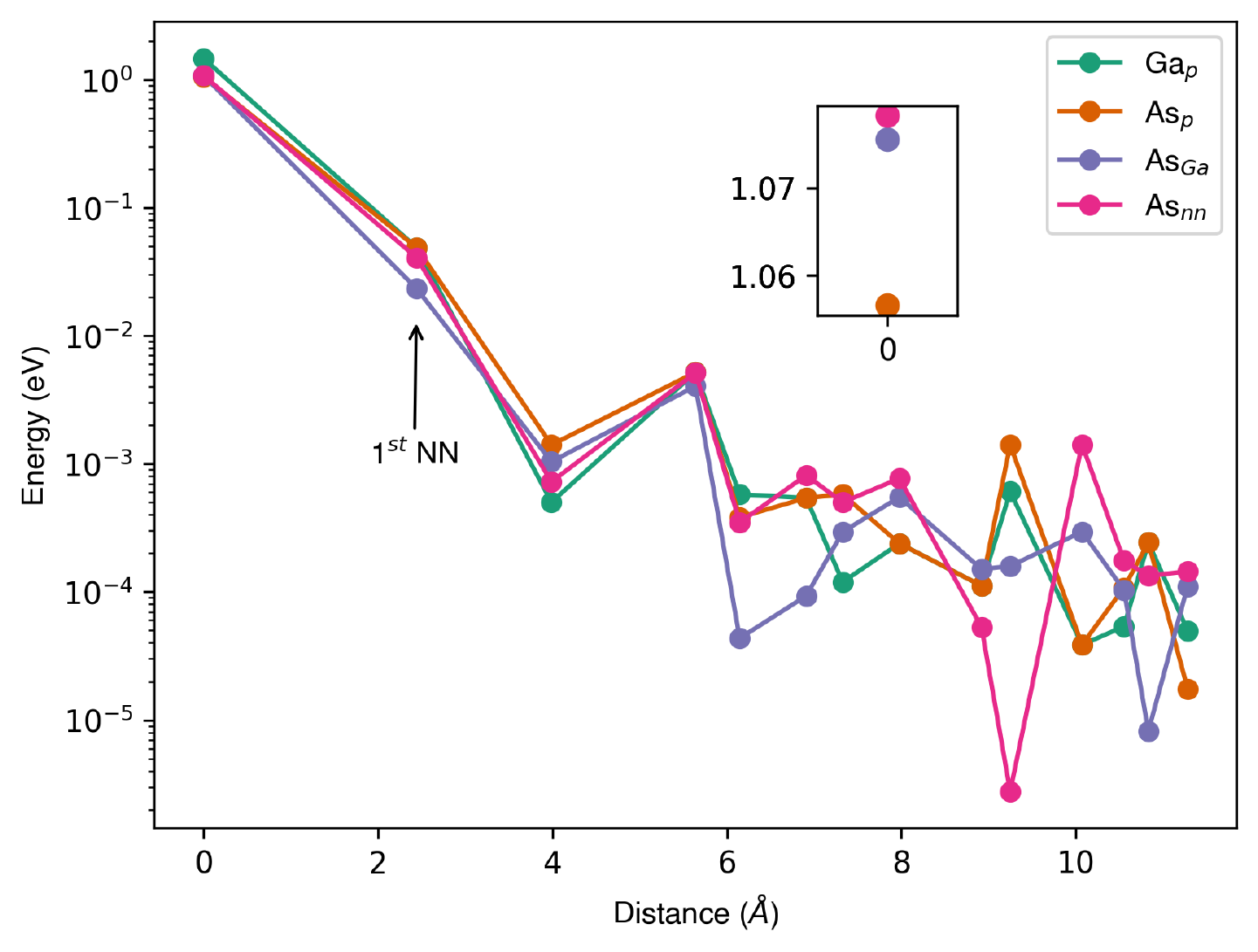}
      \caption{The trace of submatrices of corresponding atom-atom pair
               $\Tr[\Sigma_{{\bf R}_Si;{\bf R}_S'+{\bf T}_Si'}]$ in the self-energy matrix is
        plotted as function of  distance $|{\bf R}_S'+{\bf T}_S-{\bf R}_S|$.
        The basis atoms of primitive GaAs crystal
               (Ga$_p$ and As$_p$), the defect (As$_{Ga}$) from the 64 atom cell and its
               nearest neighbor (As$_{nn}$) are chosen to illustrate the drastic drop
               in energy corrections with distance. Both structures are in $q=0$ state.
               In the inset, the onsite energy corrections of the As atoms are seen
               more clearly.\label{fig:selfvsd}}
    \end{figure}

  The short-distance nature of the self-energy in real space is illustrated in
  Fig~\ref{fig:selfvsd}. We here show the trace of each block of the self-energy matrix
  for pairs $\{{\bf R}_S,{\bf R}^\prime_S+{\bf T}_S\}$ as function of the their separation
  distance. While the self-energy operator $\tilde\Sigma({\bf r},{\bf r}^\prime)$  itself
  in principle falls off as $1/|{\bf r}-{\bf r}^\prime|$, being a screened  exchange type
  term, the matrix elements
  $\Delta\tilde\Sigma_{{\bf R}_Si;{\bf R}^\prime_S+{\bf T}_Si^\prime}$ fall off much
  faster because they are dominated by the overlap of the corresponding basis function
  orbitals. The on-site elements are clearly seen to be 2-3 orders of magnitude larger
  than the inter-site elements with nearest and second nearest neighbors. They oscillate
  somewhat as we go to further neighbors and their localization could be further improved
  by means of more localized screened muffin-tin-orbitals
  such as the jigsaw orbitals proposed in
  Ref. \onlinecite{questaalpaper}. We may also notice that the on-site self-energy of As
  in the perfect crystal or in the defect site are very close to each other. The essential
  point in correcting the self-energy matrix of the perfect crystal represented in the
  supercell, which already incorporates the host band gap change, is to replace that of
  the Ga atom by an As atom. The inter-site elements of the self-energy are so much
  smaller that they play only a minor role and this explains why we can restrict the range
  of the self-energy matrix elements rather severely with a small $d_{max}$. This provides
  {\sl a-posteriori} also an explanation for why schemes such as LDA+U
  \cite{Paudel08VO,Boonchun2011} for modifying the band structure or other local on-site
  corrections \cite{Christensen84,LanyNLEP} have had considerable success in adjusting the
  gap in defect calculations.

    \begin{table}[h]
      \caption{Convergence of band gap with $d_{max}$ in GaAs with Basis set Ga: $spd$,
               As: $spd$. $a=10.66$ Bohr is the lattice constant in the zinc blende
               structure. The gap of GaAs the LDA gap with this basis set is 0.50 eV and
               the k-space QS$GW$ gap is 2.35 eV.\label{tabgap}}
        \begin{ruledtabular}
          \begin{tabular}{lrc}
            $d_{max}/a$ & $\# neighbors$ & gap (eV) \\\hline
            0.6         & 5              & 1.84     \\
            0.8         & 17             & 1.76     \\
            1.0         & 29             & 2.09     \\
            1.2         & 47             & 2.26     \\
            1.4         & 87             & 2.30     \\
            1.6         & 147            & 2.29     \\
            1.8         & 191            & 2.30     \\
            2.0         & 275            & 2.32     \\
            2.2         & 345            & 2.32     \\
            2.4         & 417            & 2.33     \\
            2.6         & 457            & 2.34     \\
            3.0         & 461            & 2.35     \\
            $\infty$    &                & 2.35
          \end{tabular}
        \end{ruledtabular}
    \end{table}

\subsection{Basic properties of the As$_{Ga}$ antisite defect} \label{asga}
  Next, we check the viability of the scheme for the case of the As$_\mathrm{Ga}$ antisite
  in GaAs. This is a well-studied defect, known as the EL2 defect, or at least closely
  related to it.\cite{Bachelet83,Dabrowski} In the $q=0$ state, it has a single $a_1$
  defect level filled in the gap and a single $t_2$ empty resonance just above the CBM.
  The excited state $a_1^1t_2^1$ is two-fold degenerate and both its
  $S=1$ and $S=0$ configurations are 
  orthogonal to the ground
  state $a_1^2t_2^0$. This degeneracy leads to a symmetry breaking distortion, such that
  the antisite As atom is pushed through the interstitial position and the initial single
  point defect turns into Ga-vacancy and As-interstitial $V_{Ga}+As_i$ defect complex.
  Due to this displacement the system might get trapped in a metastable state
  at lower energy than the excited state but at the distorted geometry.
  This metastable state is labeled $1a^02a^2$ where now the levels are labeled according to
  the $C_{3v}$ distorted geometry. The ground state 
  in this notation is $1a^22a^0$ and the excited
  state is $1a^12a^1$. In other words, $1a$ in $C_{3v}$ derives
  from the $a_1$ in $T_d$ while $2a$ derives from $t_2$.\cite{Dabrowski}
  This state is associated with the
  photoquenching behavior of EL2 defect in Ref. \onlinecite{Dabrowski}. Therefore, the
  optical excitation energy from the $a_1$ to the $t_2$ level is of interest and can be
  directly related to the corresponding QS$GW$ levels since it occurs from the ground state
  $q=0$ geometry. Furthermore, the Green's function calculations of Bachelet
  \etal\cite{Bachelet83} provide detailed information on the position of the Kohn-Sham
  defect levels of $a_1$ and $t_2$ symmetry in two charge states. We identify these two
  defect levels in defect supercell band structure and compare our results with Ref.
  \onlinecite{Bachelet83} and \onlinecite{Dabrowski}.
  The ground state defect level of neutral EL2 is well known experimentally and
  we discuss this further in Sec. \ref{largebs}.

  \subsection{Application of the method to As$_\mathrm{Ga}$.} \label{appasga}
\begin{figure}
      \includegraphics[width=8cm]{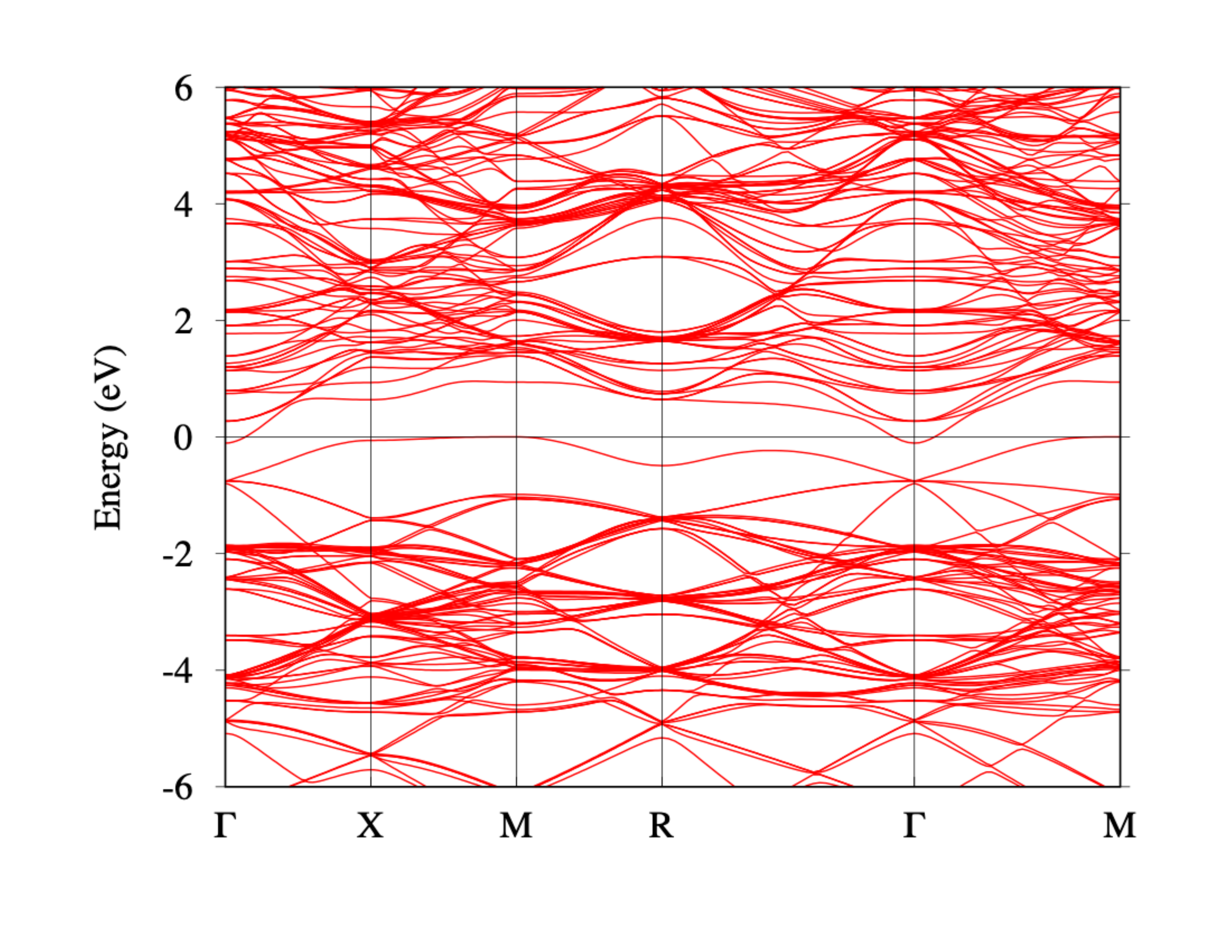}
      \caption{Band structure of As$_\mathrm{Ga}$ in $q=0$ state in GaAs in 64 atom cell
               in LDA. This supercell has simple cubic form and the high-symmetry
               ${\bf k}$-points are $\Gamma=(0,0,0)$, $X=(1,0,0)$, $M=(1,1,0)$ and
               $R=(1,1,1)$ in units $2\pi/a_S$ with $a_S=2a$ the supercell lattice
               constant.\label{fig64lda}}
    \end{figure}

  In Fig.~\ref{fig64lda} we show the band structure of a 64 atom supercell containing an
  As$_\mathrm{Ga}$ antisite defect at its origin in the LDA and for $q=0$ state. In this
  study, we used a minimal basis set of only a single $\kappa$ and smoothing radius
  $R_{sm}$, $spd$ orbital set on Ga and As. The gap is thereby underestimated as 0.65 eV
  in LDA and overestimated as 2.25 eV in QS$GW$.  This should facilitate recognizing the
  defect level. Nonetheless, we see in Fig.~\ref{fig64lda} that the electronic structure
  of the defect and even the gap are barely recognizable. The defect band is the highest
  occupied band but is seen to be so much broadened that, in combination with the LDA gap
  underestimate, it touches the valence band maximum (VBM) at the $\Gamma$-point. The
  conduction band minimum (CBM) occurs 0.65 eV above it at the $\Gamma$-point in
  LDA, while a converged basis set would give an even lower LDA gap of only 0.51 eV.
  Clearly, the band structure of this system cannot be examined accurately with DFT level
  calculations.

\begin{figure}
      \includegraphics[width=8cm]{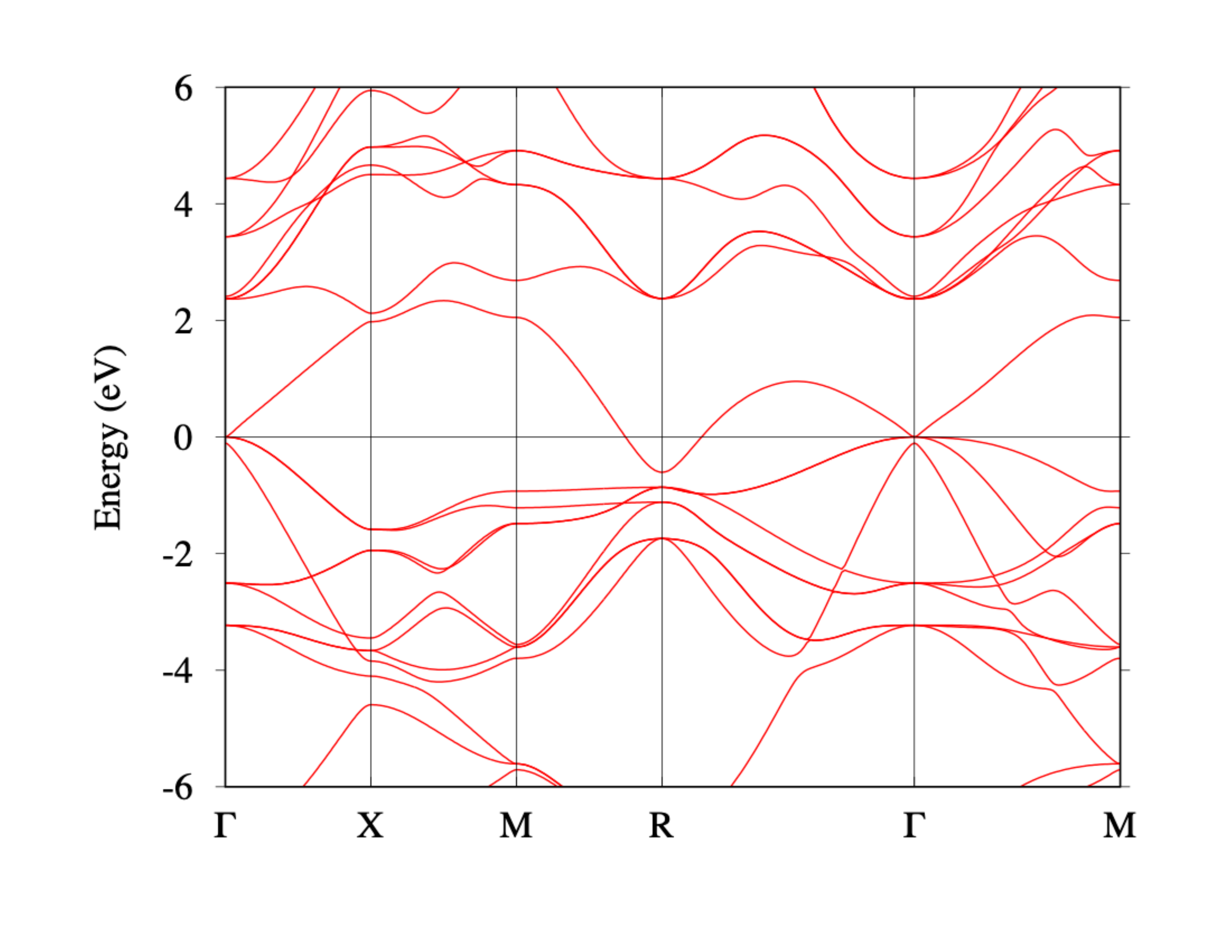}
      \caption{Band structure of As$_\mathrm{Ga}$ in 8 atom GaAs cell in QS$GW$. This
               supercell is also a simple cubic but with lattice constant $a$. The
               high-symmetry points are labeled the same as in Fig. \ref{fig64lda} but
               correspond to a Brillouin zone twice as large in each direction.
               \label{fig8gw}}
    \end{figure}

  Next, we construct the defect in an 8 atom cell and perform a QS$GW$ calculation for it.
  From the corresponding band structure shown in Fig.~\ref{fig8gw} it is clear that this
  cell is much too small to adequately represent the defect electronic structure. In this
  figure, it is not even clear which is the defect band and which is the lowest conduction
  band. On the other hand, we will show that this cell is sufficient to reconstruct the
  self-energy components in the immediate neighborhood of the defect.

\begin{figure}
      \includegraphics[width=8cm]{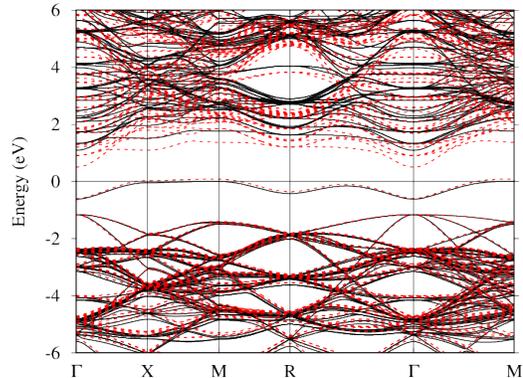}
      \caption{Band structure of As$_\mathrm{Ga}$ in $q=0$ state in GaAs in 64 atom cell:
               fully QS$GW$ (solid black lines), with self-energy constructed by
               cut-and-paste approach (red dashed lines) using only the defect atom as
               defect region aligned at the VBM. The zero of energy is the Fermi level for
               the defect system in the full QS$GW$ case for the neutral defect.
               \label{fig64gw}}
    \end{figure}

  In Fig.~\ref{fig64gw}, we show the band structure of the 64 atom cell obtained by means
  of our self-energy cut-and-paste approach with the dashed red lines. The defect region
  contains only the defect atom itself and the self-energy range $d_{max}$ was set to one
  lattice constant (5.64 \AA). In the same figure we show the fully self-consistent QS$GW$
  results in the same 64 atom cell with the solid black lines, after aligning the two at
  the VBM. The dispersion of this defect band is about 0.7 eV and results from
  periodically repeated defects in the 64 atom supercells, so from defects that are
  2$a_{cell}$ or 11.28 \AA\ apart. Comparing with the band structure in a larger cell in
  Fig. \ref{fig:216} we can see that the defect band width is mostly reduced at $R$ and
  $\Gamma$ but the top of the band near $X$ and $M$ stays the same. We therefore identify
  the eigenvalue near its maximum with the isolated defect level. The fact that the defect
  band  dispersion does not follow the expected form
  $E_d+2t[\cos{(k_xa)}+\cos{(k_ya)}+\cos{(k_za)}]$ for a simple isotropic $s$-band in a
  nearest-neighbor simple cubic lattice, where the maximum would be at $R$ and the minimum
  at $\Gamma$ indicates that the interactions between defect states are not isotropic
  because of the underlying crystal structure. The filled defect band is now clearly
  detached from the VBM and it occurs at about 1.12 eV above the VBM. Its position above
  the VBM and even its dispersion is in excellent agreement between the cut-and-paste and
  fully self-consistent results. Furthermore, this defect level position above the VBM is
  in good agreement with Bachelet's Green's function calculation\cite{Bachelet83} of an
  isolated defect, which gives 1.23 eV for the $q=0$ state.

  The CBM is also clearly seen (overestimated as 2.25 eV above the VBM by our small basis
  set) in the full self-consistent calculation. This gap is somewhat smaller in the
  cut-and-paste case (1.66 eV). This is because the range cutoff $d_{max}$ applied to
  $\Sigma$ reduces the gap. Second, in this 64 atom cell, we do not yet approach the
  dilute limit where the perfect crystal gap with this cutoff (2.09 eV) would be
  recovered. This can be attributed to the $t_2$ defect level interacting with the CBM. In
  both cases, we can identify the second defect level, the $t_2$ state, which here is a
  resonance and can be recognized as a flat band near $X$ at about 1.55 eV above the Fermi
  level which coincides with the top of the $a_1$ defect band, in direct QS$GW$
  calculation and 1.07 eV in the cut-and-paste method. This defect level lies 0.42 eV
  above the CBM in direct QS$GW$ calculation and 0.60 eV in the cut-and-paste approach.
  The energy splitting between the $t_2$ and $a_1$ state taken as the energy difference at
  $X$-point, because of above reasons, hence 1.55 eV in direct calculation and 1.06 eV in
  our approach. Dabrowski \etal\cite{Dabrowski} reports this splitting as 1.18 eV
  experimentally and as 0.97 eV at the DFT level calculations, while in the Green's
  function calculation it is 0.87 eV \cite{Bachelet83}. Our result for this splitting of
  the $t_2-a_1$ level is thus comparable in accuracy with the previous calculations and
  in fact closer to the experimental value, which, as mentioned earlier, is important for
  understanding the optical behavior of this defect. Hence, the the cut-and-paste method
  is found to be a viable approach.

  Next, we test whether replacing only the self-energy related to the defect atom itself
  is sufficient or whether we need to include a larger defect cluster region. In our
  cut-and-paste method, including the nearest neighbors in the defect region corresponds
  to taking the nearest neighbor of the defect atom as another center for the
  $\Delta\tilde\Sigma_{{\bf R}_S,{\bf R}^\prime_S+{\bf T}_S}$ to be taken from the small
  defect containing (8 atom) cell. This approach could be problematic, because it extends
  the range of this $\Delta\tilde\Sigma$ with ${\bf R}_S$ being a nearest neighbor of the
  defect beyond its unit cell, hence includes pairs connecting this atom to other defect
  atoms, while in the dilute limit, or in the large supercell with a single defect, there
  should only be one defect atom. However, because the off-site matrix elements fall off
  rather quickly, this is found not to be a serious problem. With this approach, we
  observe that the valence and defect bands remain in the same position, but the CBM value
  drop 0.14 eV, which deviates from the QS$GW$ calculation as expected. These results are
  shown in supplemental information.

    \begin{figure}
      \includegraphics[width=8cm]{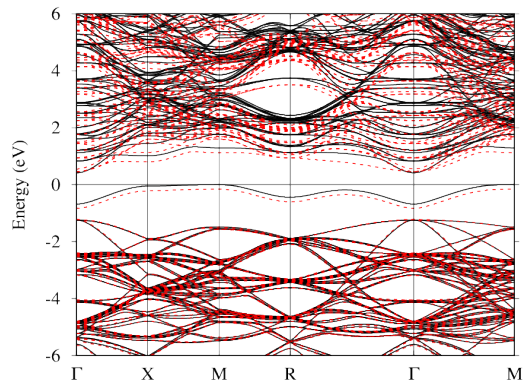}
      \caption{Band structure of As$_\mathrm{Ga}$ in $q=0$ state in GaAs in 64 atom cell:
               $64_8$ scheme (solid black lines), and $64_{32}$ scheme (red dashed lines),
               both with self-energy constructed by cut-and-paste approach using only the
               defect atom as defect region, aligned at the VBM. The zero of energy is the
               Fermi level for the defect system in $64_8$ scheme.\label{8to32}}
    \end{figure}

  Clearly, in order to allow us to include more neighbors than the defect atom itself in a
  safe way, we would need to enlarge the size of the cell from which the defect and its
  neighbors self-energy is extracted. This might then also allow to increase the
  $d_{max}$ or range of the self-energy cutoff. To further test the convergence of our
  scheme we now consider a somewhat larger cell than the 8 atom cell to extract the defect
  atom and its neighbors' self-energy. For this purpose, we incorporate the defect in a 32
  atom supercell of GaAs. The $d_{max}$ value is set to the lattice constant of this cell,
  1.73$a$ of the conventional cell, which is then also the nearest defect distance.
  
  Building the host cell using this small cell does not provide considerable improvement
  for the cut-and-paste method. This is expected since we already know that the
  self-energy matrix elements fall off rapidly with intersite distance. Furthermore, the
  exact QS$GW$ calculation of this cell is now a more expensive calculation and not so
  much is gained by increasing the size to a 64 atom cell using the cut-and-paste
  approach. Nonetheless, we test its performance to check the convergence of our
  cut-and-paste approach in terms of the size of the defect region and self-energy cutoff
  distance $d_{max}$. The results of this scheme is summarized in
  Table~\ref{tablesumsmall} with the label 64$_{32}$. Different defect region descriptions
  and charge states were also considered. In Fig.~\ref{8to32}, we compare the
  two cut-and-paste band structures of the defect containing 64 atom cell in the $q=0$
  state, built from the 8 atom cell (64$_{8}$) and from the 32 atom cell (64$_{32}$). The
  almost identical bands show that there is nothing to be gained in the accuracy of the
  method by extracting the defect atom self-energy from a 32-atom compared to an 8 atom
  cell and is furthermore significantly more computationally expensive. We conclude that
  the most effective approach for the cut-and-paste method is building the defect
  containing host cell from the 8 atom cell and it is sufficient to define the defect
  region as the defect atom itself. The results obtained by this approach are in good
  agreement with our exact QS$GW$ calculations and previous studies in the literature.

    \begin{figure}
      \includegraphics[width=8cm]{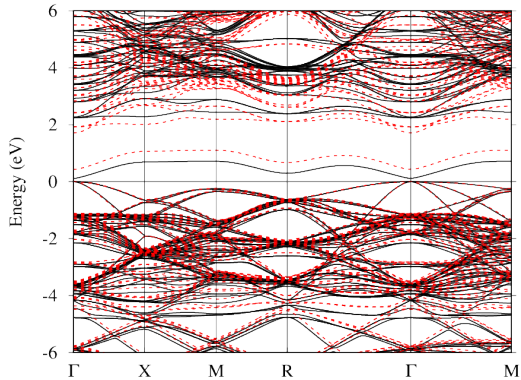}
      \caption{Band structure of As$_\mathrm{Ga}$ in $q=2$ state in GaAs in 64 atom cell:
               fully QS$GW$ (solid black lines), with self-energy constructed by
               cut-and-paste approach (red dashed lines) using only the defect atom as
               defect region, aligned at the VBM. The zero of energy is the Fermi level
               for the VBM in the full QS$GW$ case for the doubly ionized defect.
               \label{fig64gw+2}}
    \end{figure}

  Next we want to address whether the cut-and-paste approach works also for charged defect
  states and correctly describes the trend with charge observed in the Green's function
  calculations. In Fig.~\ref{fig64gw+2} we show the band structure of the same cell, but
  with the defect in $q=2$ state. In this case, the exact QS$GW$ band gap does not change,
  but the defect level moves closer to the VBM. The cut-and-paste approach yields 0.05 eV
  larger band gap. Although the defect band dispersion is again reproduced faithfully, the
  position of defect level deviates by about 0.2 eV between the cut-and-paste and exact
  calculation. This is still an acceptable precision of the cut-and-paste method and
  indicates that the nearly perfect agreement seen earlier for the $q=0$ state is perhaps
  somewhat coincidental.

  On the other hand, the difference between the two charge states regarding the defect
  level position agrees with prior work and has a clear physical meaning. It can be
  related to the different final geometries after atomic relaxations in different charge
  states. For the $q=0$ state, we observe outward breathing of nearest As atoms, such that
  they move further from the defect atom about 4\% of the nearest neighbor distance.
  Dabrowski \etal\cite{Dabrowski} reported similar lattice relaxation effects.
  On the other hand, in the $q=2$ state, the  nearest neighbor
  distance remains the same as in the perfect crystal.

    \begin{figure}
      \includegraphics[width=8cm]{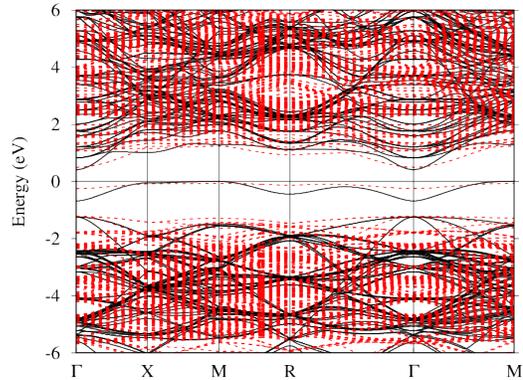}
      \caption{Band structure of As$_\mathrm{Ga}$ in $q=0$ state: $64_8$ scheme (solid
               black lines), and $216_8$ scheme (red dashed lines), aligned at the VBM.
               Defect region is defined only with the defect atom. Dispersion of $a_1$
               level vanishes as defect-defect distance increases, in a fashion explained
               with the tight-binding model above. Note that the Brillouin zones of the
               216 and 64 atom cells have the same symmetry labeling but the latter is
               larger and contains fewer bands. The $\Gamma-X$ distance of the 216 atom
               cell is scaled down by a factor 2/3 compared with the 64 atom one and
               similarly for the other directions.\label{fig:216}}
    \end{figure}

  One important advantage of our method is that once the exact QS$GW$ calculation for the
  small cell is done, moving to larger host cells and calculating the defect properties in
  the dilute limit is a straightforward procedure and one can achieve $GW$-level
  accuracies for the cost of an LDA calculation for even larger cells. Although we do not
  include the exact calculations for these larger cells in this work, we analyze a 216
  atom cell, \ie a $3\times3\times3$ supercell of the conventional simple cubic 8 atom
  cell. In Fig.~\ref{fig:216} we see that in the 216 atom cell dispersion of the defect
  band is almost zero, indicating we are close to the dilute limit. Furthermore, this
  allows us to better evaluate the nature of the defect band dispersion. As mentioned
  earlier, the top of the defect band which is flat between $X$ and $M$ remains the same
  as in the smaller 64 and 32 atom cells but the band width is reduced at $\Gamma$. This
  helps us to identify the top of the defect band with the dilute limit defect level.

    \begin{table}[h!]
      \caption{The energy band gap, $E_g$, the defect level at the $X$-point
        w.r.t.\ VBM,  $a^X_{1_{VBM}}$,  the defect level splitting at the $X$-point,
        $(t_2-a_1)^X$ and
               the $a_1$ defect band width $w(a_1)$, are presented for various schemes in
               $q=0$ (top half) and in $q=2$ (bottom half) states. The subscript numbers
               represent the small cell atom number which was used to build up the final
               cell in the cut-and-paste method.\label{tablesumsmall}}
        \begin{tabular}{lcccc}
                                                & $E_g$     & $a^X_{1_{VBM}}$ & $(t_2-a_1)^X$ & $w(a_1)$\\\hline

          Dabrowski\cite{Dabrowski}                             &           & 0.6             & 0.97                &              \\
          Bachelet\cite{Bachelet83} (LDA)                       & 0.7  & 1.23            & 0.87                &              \\
          64 atom QS$GW$  & 2.25      & 1.12            & 1.55                & 0.62         \\\hline
          64$^{\dagger}_8$                      & 1.66      & 1.19            & 1.07                & 0.68         \\
          64$^{\ast}_8$                         & 1.52      & 1.19            & 1.04                & 0.69         \\
          64$^{\dagger}_{32}$                   & 1.69      & 1.01            & 1.02                & 0.67         \\
          64$^{\ast}_{32}$                      & 1.64      & 0.98            & 1.03                & 0.68         \\
          216$^{\dagger}_{8}$                   & 1.72      & 1.14            & 1.24                & 0.19         \\
          216$^{\ast}_{8}$                      & 1.67      & 1.14            & 1.23                & 0.19         \\\hline\hline
          Bachelet                              &           & 0.69            & 0.99                &              \\
          64 atom QS$GW$                      & 2.24      & 0.69            & 1.60                & 0.60         \\\hline
          64$^{\dagger}_8$                      & 1.71      & 1.06            & 0.95                & 0.67         \\
          64$^{\ast}_8$                         & 1.58      & 1.07            & 0.94                & 0.67         \\
          64$^{\dagger}_{32}$                   & 1.63      & 0.92            & 0.96                & 0.67         \\
          64$^{\ast}_{32}$                      & 1.64      & 0.92            & 0.95                & 0.67         \\
          216$^{\dagger}_{8}$                   & 1.70      & 0.61            & 1.24                & 0.16         \\
          216$^{\ast}_{8}$                      & 1.65      & 0.62            & 1.24                & 0.16         \\\hline

        \end{tabular}
      \begin{tablenotes}
        \small
          \item $^{\dagger}$ Defect region is defect atom itself.
          \item $^{\ast}$ Defect region is defect atom and its nearest neighbors.
      \end{tablenotes}
    \end{table}

  It is important to point that obtaining this 216 atom cell band structure only required
  an additional LDA calculation of this cell. Once the 8 atom cell $GW$ calculation is
  done, moving to even larger cells only requires the LDA calculation of the desired
  single defect containing supercell. This provides an immense reduction in computational
  cost. For instance, we can compare the computation time of exact QS$GW$ calculations for
  8 atom, 32 atom and 64 atom cells. The convergence parameters for these calculations are
  kept the same and the {\bf k}-point meshes are kept equivalent by taking a smaller set
  with approximately the same grid-density in the larger cells. The number of processors
  for each calculation is adjusted proportional to the {\bf k}-point mesh grid. Under
  these circumstances, fully consistent QS$GW$ calculation was completed in 0.6 hour for
  the 8 atom cell, 26.4 hour for the 32 atom cell, and 181 hours for the 64 atom cell. In
  other words, the computing time scales like $N^\eta$ with $\eta\approx1.8$ and $N$ the
  number of atoms or $3\eta$ for the scaling with the linear size of the supercell. On the
  other hand, the LDA calculation of the 64 atom cell took less than half an hour. For
  this particular 64$_8$ scheme, the bottleneck of cut-and-paste method was the 8 atom
  cell QS$GW$ calculation. For larger single defect cells, the LDA calculation might
  become the bottleneck due to large number of atoms, however the computation cost would
  still be insignificant, compared to the QS$GW$ calculation of the same cell. For
  instance, the LDA calculation of the host cell in our 216$_{8}$ scheme was performed
  using resources comparable to that of 64$_8$ LDA step, and this step took around 6
  hours. Note that all the atoms are fully relaxed in this step, and one could reduce the
  computation time even further by only letting the first few nearest neighboring atoms
  relax.

\subsection{Large basis sets and comparison to experiment} \label{largebs}
To test the fidelity of the theory, we finally make a careful comparison against the experimentally observed neutral EL2 deep donor 
level ($E_{v}{+}0.75$\,eV).\cite{Dabrowski,Weber82} This defect level
was first identified with the As$_\mathrm{Ga}$ antisite by Weber \etal\cite{Weber82} based on Electron
Paramagnetic Resonance (EPR) and activation of the unpaired spin $+1$
charge state from the neutral state by an optical transition
to the conduction band, which was found to occur at 0.75 eV. This was later
also confirmed by thermionic emission and optical studies.\cite{Omling86,Samuelson86}
In view of the experimental gap of 1.52 eV at low temperature,
this places the defect level almost exactly at mid gap.
We revisit
the neutral EL2 level with a reasonably well converged basis, and also include spin-orbit coupling (SOC).  We repeat the
procedure described above, using a 32-atom supercell, with an \emph{spdfspd} basis on both the Ga and As, and local
orbitals to include the Ga $3d$ in the valence.  The basis also includes \emph{sp} ``floating orbitals'' (smoothed
Hankel functions without augmentation spheres~\cite{questaalpaper}) centered at the two high-symmetry interstitial sites
along the [111] line.  The QS\emph{GW} bandgap (1.7\,eV, including SOC) is slightly less than that of a fully converged
basis (1.8\,eV).  This includes a reduction in the bandgap of 0.11\,eV from SOC.  The gap reduction is expected, as the
split-off valence band at $\Gamma$ is 0.33\,eV below the VBM, both experimentally and in QS\emph{GW}.  

We note that even with his well-converged basis set the bandgap is overestimated because QS\emph{GW} under-screens $W$, because the RPA polarizability omits electron-hole
attractions that connect the electron and hole parts of the bubble.  If such attractions are included, \eg via ladder
diagrams, the bandgap is reduced to a value very close to the observed gap.  It was discovered soon after QS\emph{GW}
was first formulated that, empirically, the RPA dielectric constant $\epsilon_\infty$ is uniformly 20\% too small for a
wide range of insulators, strongly correlated or not~\cite{Chantis06a}; see in particular Fig. 1 in
Ref.~\cite{Bhandari18}.  Adding ladder diagrams greatly improves on $\epsilon(\omega)$, as has been known from
pioneering work in the groups of Louie\,\cite{Rohlfing98b} and Reining\,\cite{Albrecht98}.  It has recently shown that
improving $W$ with ladders, and using this $W$ in the QS\emph{GW} self-consistency cycle, almost completely eliminates
the overestimate of bandgaps in weakly correlated semiconductors, and it also corrects for the underestimate of
$\epsilon_\infty$ \cite{cunningham21}.  As an alternative, a hybrid approach has long been used, mixing LDA and
QS\emph{GW}~\cite{Chantis06a}. Kotani and his coworkers showed that a hybrid of 80\% QS\emph{GW} and 20\% LDA yields
uniformly good bandgaps in many weakly correlated semiconductors ~\cite{Deguchi16}.
The hybrid approach is an inexpensive, albeit \emph{ad hoc} way to mimic the
effect of ladder diagrams, and we use it here to refine our estimate of the neutral EL2 level.

\begin{figure}
  \includegraphics[width=7cm]{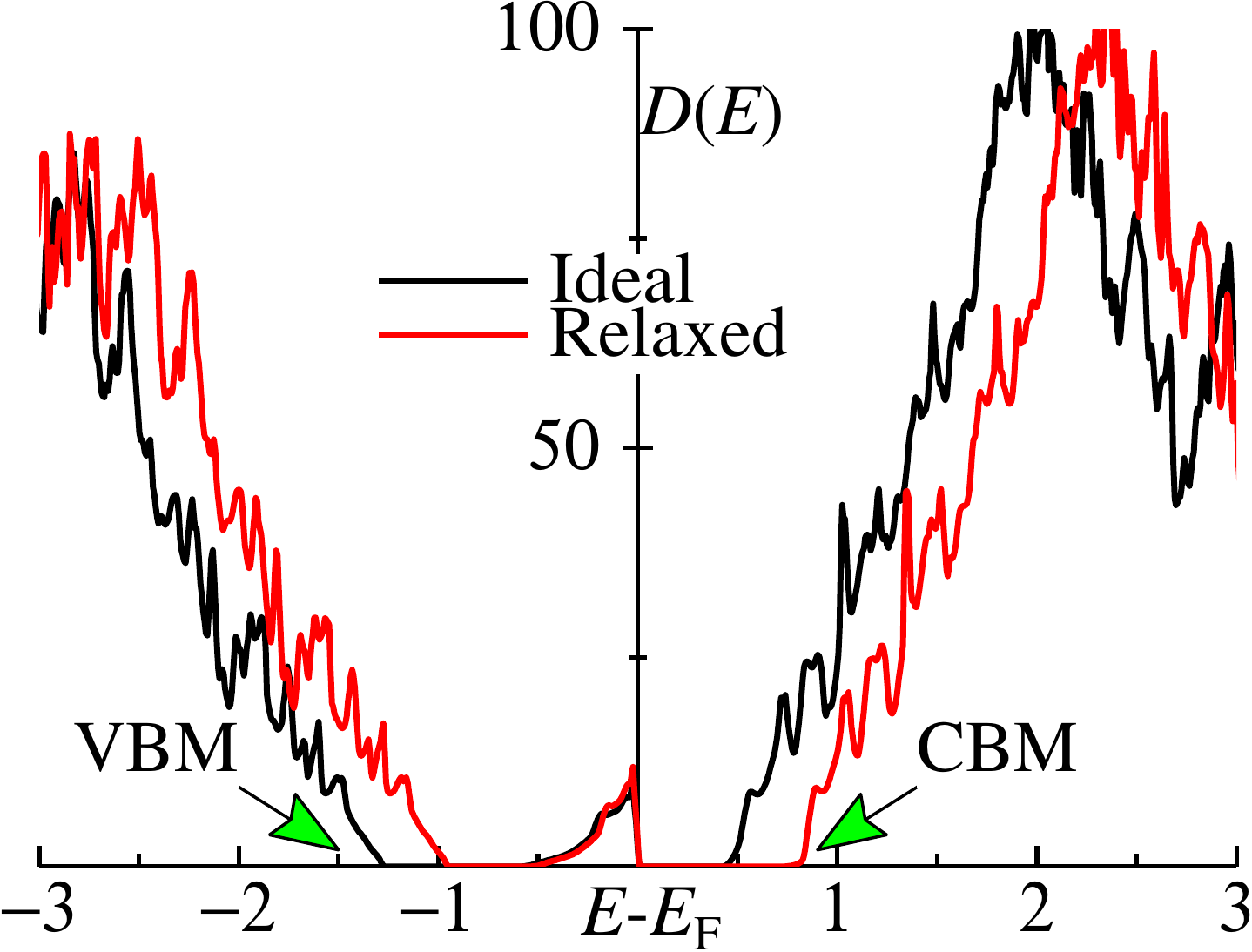}
  \caption{Density-of states of a 128 atom supercell of GaAs with a single As antisite for an unrelaxed lattice (black),
    and for a lattice whose nearest neighbors were relaxed (red).  Energy zero corresponds to the Fermi level in both
    cases, and sits at the top of the (doubly occupied) midgap EL2 level. It has a dispersion of about 0.3\,eV.  
    in the 128 atom cell.
  \label{figdos}}
\end{figure}

Fig.\,\ref{figdos} shows the density-of-states (DOS) of the 128 supercell within QS\emph{GW} for two scenarios: the black data
is the DOS for an ideal (unrelaxed) structure, while the red data shows the effect of relaxing the four nearest
neighbors around the As\textsubscript{Ga} only.  Limiting relaxations to nearest neighbors simplifies the embedding
procedure, and tests showed that more complete relaxations made minor further changes.  It is seen that the lattice
relaxation induces a shift in the defect level, moving it about 0.3\,eV closer to the valence band.

To estimate the EL2 energy in the limit of an infinite cell,
the center of gravity of the band was calculated.  It is found to be 
close to the Fermi level at the top of the defect band for the
neutral charge state, about 0.11 eV below $E_F$. Thus, for higher precision
we here subtract this 0.11 eV from $E_F$ to obtain the defect level.
This would reduce the $a_1$ defect levels in Table  \ref{tablesumsmall}
by 0.11 eV. 
The VBM and CBM are inferred from the
energies where the DOS touches zero (see labels in Fig.\,\ref{figdos}), so that the EL2 relative to either can be
computed.  The results are displayed in Table~\ref{tabbigel2}.

\begin{table}[h]
  \caption{Band center of neutral EL2 defect, in eV, relative to the valence band maximum, embedded in a 128 atom cell.
  A reasonably well converged basis was used, with spin orbit coupling included.  Two calculations are shown:
  the first assuming no lattice relaxation, and the second including it, as described in the text.
  Also shown is the bandgap. \label{tabbigel2}}
    \begin{ruledtabular}
      \begin{tabular}{ccccc}
                    & \multispan2 QSGW  & \multispan2 80\%QSGW+20\%LDA \cr
        structure   & EL2  & gap     &\qquad EL2  & gap  \cr
        unrelaxed   & 1.13 & 1.69    &\qquad 1.02 & 1.44 \\
        relaxed     & 0.83 & 1.68    &\qquad 0.74 & 1.42 
      \end{tabular}
    \end{ruledtabular}
\end{table}

As Table~\ref{tabbigel2} shows, EL2 is predicted to be at $E_{v}{+}0.83$\,eV and $E_{c}{-}0.85$\,eV.  It is slightly too
far from both the VBM and the CBM, compared to experiment, because the QS\emph{GW} gap is too large.  It is not \emph{a priori} obvious how much this level will shift if QS\emph{GW} were high enough fidelity to yield the experimental gap,
\eg by adding ladder diagrams to $W$.  It is possible in principle to do this by carrying out the calculation with
ladders in $W$, but here we take the hybrid 80\%QS\emph{GW}+20\%LDA approach as a simple alternative.  The result is shown in
Table~\ref{tabbigel2}.  The hybrid approach reduces the level by about 0.1\,eV, while the gap itself is reduced by
~$\sim$0.25\,eV.  This is consistent with the EL2 being comprised of roughly equal measures of the host valence band and
conduction band character and its position almost exactly in the middle of the gap.  Finally, the predicted EL2 energy ($E_{v}{+}0.74$\,eV) is in excellent agreement with the
observed value $E_{v}{+}0.75$\,eV, while the bandgap (1.42\,eV) is also close to the room-temperature experimental gap.
(Ideally the QS\emph{GW} gap should be $\sim$1.5\,eV, the zero-temperature gap of GaAs.  Indeed, it does come out very
close to 1.5\,eV when a fully converged basis is used, whether ladders are added or the hybrid-$\Sigma$ approach is
used.)

Even with a slightly less complete basis set, including $spdfspd$ and
Ga-$3d$ local orbitals, but omitting the floating orbitals, we found
that the defect band top in a 64 atom cell and using $64_8$ cut-and-paste
approach, lies at 0.78 eV above the VBM when the latter is corrected by SOC,
and with a host
gap of 1.7 eV. Given that the center of gravity of the defect band DOS lies
slightly lower, this would become 0.67 eV.
Even in this calculation, however, the conduction
band minimum in the defect cell is $\sim$1.35 eV, indicating that the
$t_2$ resonance in the conduction band  affects the conduction band minimum.
This effect is reduced only by going to even larger supercells. 

\section{Conclusion}
  We have shown that the $GW$ self-energy matrix represented in a real-space basis-set is
  short-ranged, and the contribution to the quasiparticle energy correction to DFT
  eigenvalues is significant only for the first few nearest neighbors of a specific atom.
  We introduce a cut-and-paste method for defect calculations at the $GW$ level that
  exploits this property. We demonstrate the method using a well-known single point
  defect, namely the As$_{Ga}$ antisite in a GaAs crystal. The main correction to the
  defect band structure compared to LDA in our method is incorporating the host perfect
  crystal gap correction via the perfect crystal self-energy, which requires a trivial
  cost because it just amounts to a relabeling of the self-energy matrix according to the
  supercell description of the atomic sites. After detailed examination, we conclude that
  an 8 atom cell is sufficient to extract the defect atom's self-energy which is then used
  to replace the defect atom self-energy in the final supercell. We observe almost perfect
  agreement between the fully self-consistent QS$GW$ defect bands and the cut-and-paste
  method, in terms of the valence bands, the $a_1$ defect level position and the defect
  band dispersion. There is a small disagreement between our method and the full QS$GW$
  calculations, in terms of unoccupied levels. This is caused mainly by the range cutoff
  $d_{max}$ applied to $\Sigma$, which slightly reduces the gap from its converged QS$GW$
  value. The main advantage of the method is that it allows us to obtain $GW$-level
  accuracy results for large defect supercells at essentially the cost of an LDA
  calculation for the latter. This allowed us to carefully monitor the defect band
  dispersion and identify the dilute limit isolated defect level more precisely. We found
  good agreement for the defect level positions with previous studies of this system,
  although the previous studies were not at the $GW$-level. Inspecting the band structures
  of the defect system in detail allowed us to identify not only the obvious defect level
  in the gap of $a_1$ symmetry but also the excited defect $t_2$ symmetry resonance and
  provides accurate information on the optical transition between these levels, which has
  previously been recognized as an important step in activating a metastable state of this
  defect. Our calculation also agrees with previous work in the change in defect level as
  function of charge states. Overall, the cut-and-paste method significantly reduces the
  computational cost of $GW$-level calculations, with a small loss in accuracy.
 Finally, to establish that QS\emph{GW} is able to predict defect levels with
  a fidelity comparable to its ability to predict energy bands of bulk materials, 
  we benchmark the embedding approach against the experimentally measured neutral EL2 level. We show
  the discrepancies with experiments are small, and closely track the known discrepancies
  for weakly correlated periodic systems.

\acknowledgments{This work was supported by the US Department of Energy Basic Energy
                 Sciences (DOE-BES) under grant number DE-SC0008933. The calculations were
                 performed on the High Performance Computing Resource in the Core Facility
                 of Advanced Research Computing at Case Western Reserve University. 
                 M.v.S. was supported by the U.S. Department of Energy, Office of Science, 
                 Basic Energy Sciences, under award FWP ERW7246.}

\bibliography{gw,lmto,defect,dft}
\newpage
\appendix
\begin{widetext}
\section{Supplemental materials}
\subsection{Cut-and-paste method scheme}
  All the steps involved in the cut-and-paste method are explained here. The first step is
  obtaining the perfect cell self-energy matrix for the large supercell, which is intended
  to calculate the defect properties. Once we perform the QS$GW$ calculation for a
  primitive cell, the obtained self-energy matrix can be mapped to a larger perfect cell
  using the ``self-energy editor'' provided by the Questaal code suite. In our example, we
  started with the GaAs primitive cell (a 2 atom cell), and mapped it into two larger
  cells, namely an 8 atom cell and a 64 atom cell. Second, we introduce the As$_{Ga}$
  antisite defect to the 8 atom cell, and perform the atomic relaxation and QS$GW$
  calculations for it, subsequently. This calculation will provide the self-energy matrix
  of the immediate neighborhood of the defect. Then, this self-energy matrix also needs to
  be mapped into a 64 atom cell, to match with the matrix size of perfect cell
  self-energy. Note that the mapping steps take virtually no time because they merely
  involve a re-labeling of sites. Finally, we perform the LDA calculation for 64 atom cell
  with a single defect, which will provide the Kohn-Sham eigenvalues.

  At this point, we have all the ingredients to finalize the method. The last task before
  the ``merging'' step is to decide on the defect region. In our calculations, we show
  that describing the defect region as the defect atom itself provides accurate results.
  In the merging step, the part of self-energy matrix associated with atoms within the
  range $d_{max}$ from the defect atom in $64^{8d}$ replaces the corresponding matrix
  elements of the perfect cell self-energy matrix. The steps in the procedure is
  illustrated in the flow diagram of Fig.~\ref{fig:flow}.

    \begin{figure*}
      \includegraphics[width=12cm]{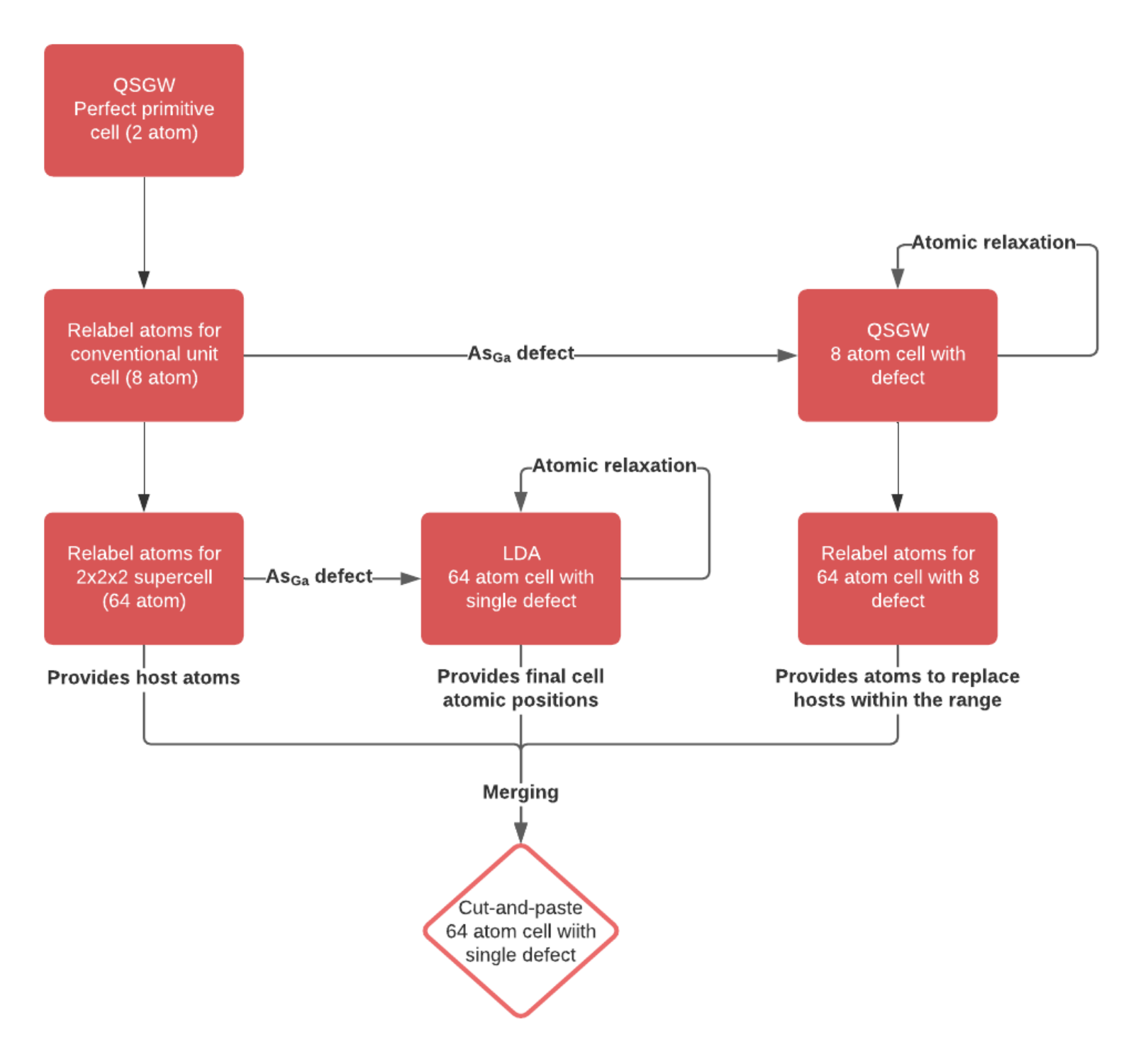}
      \caption{Schematic demonstration of cut-and-paste method.\label{fig:flow}}
    \end{figure*}

\subsection{Additional Results}
  Here we present additional information on the band structure calculations for different
  choices of the parameters in the proposed cut-and-paste scheme. In the main paper, we
  claimed that it suffices to consider the changes in self-energy from the perfect crystal
  for only the defect atom itself. Here we provide evidence of this by giving the
  corresponding results when the defect region contains the defect atom and its nearest
  neighbors. Although we do not recommend this approach because it may cause errors such
  that some longer range intersite self-energy matrix elements may incorrectly connect the
  atoms in the defect region to mirror images of the defect, we here show the results for
  various schemes (size of final and intermediate supercells) in Fig. 2, the main results
  of which were summarized and in Table-II in the paper.

  We also show the comparison of different schemes to 64$_8$ scheme in $q=2$ state, when
  the defect region is described as the defect atom (Fig. 3), and when it is described as
  the defect and its nearest neighbors (Fig.4). This illustrates that the 64$_8$ scheme is
  also sufficient in $q=2$ state. Detailed quantitative results of all calculations are
  presented in below in Table-I and Table-II.

    \begin{figure*}
      \centering
        \begin{subfigure}[b]{0.3\textwidth}
          \centering
            \includegraphics[width=\textwidth]{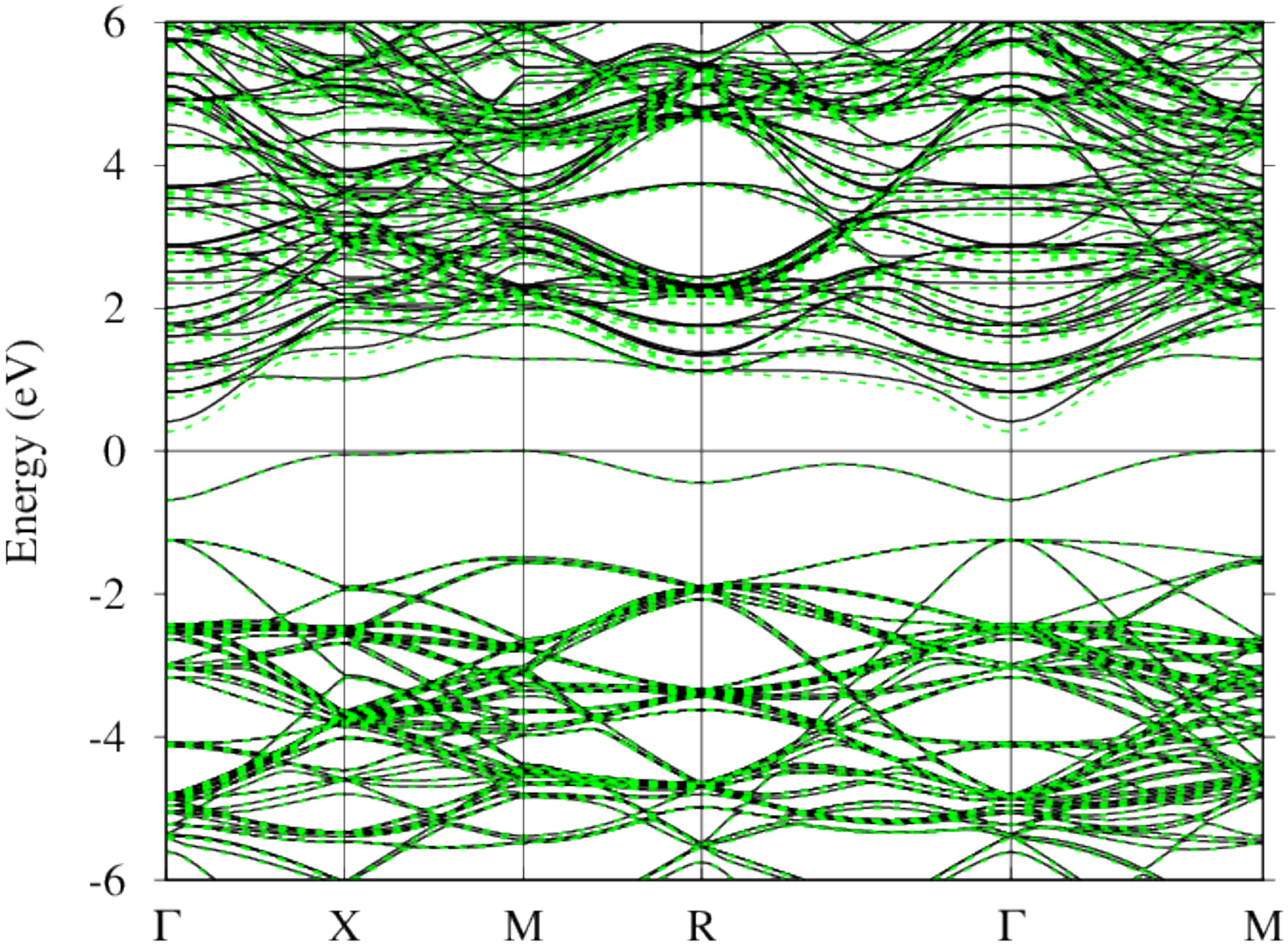}
            \caption{}
            \label{}
        \end{subfigure}
          \hfill
        \begin{subfigure}[b]{0.3\textwidth}
          \centering
            \includegraphics[width=\textwidth]{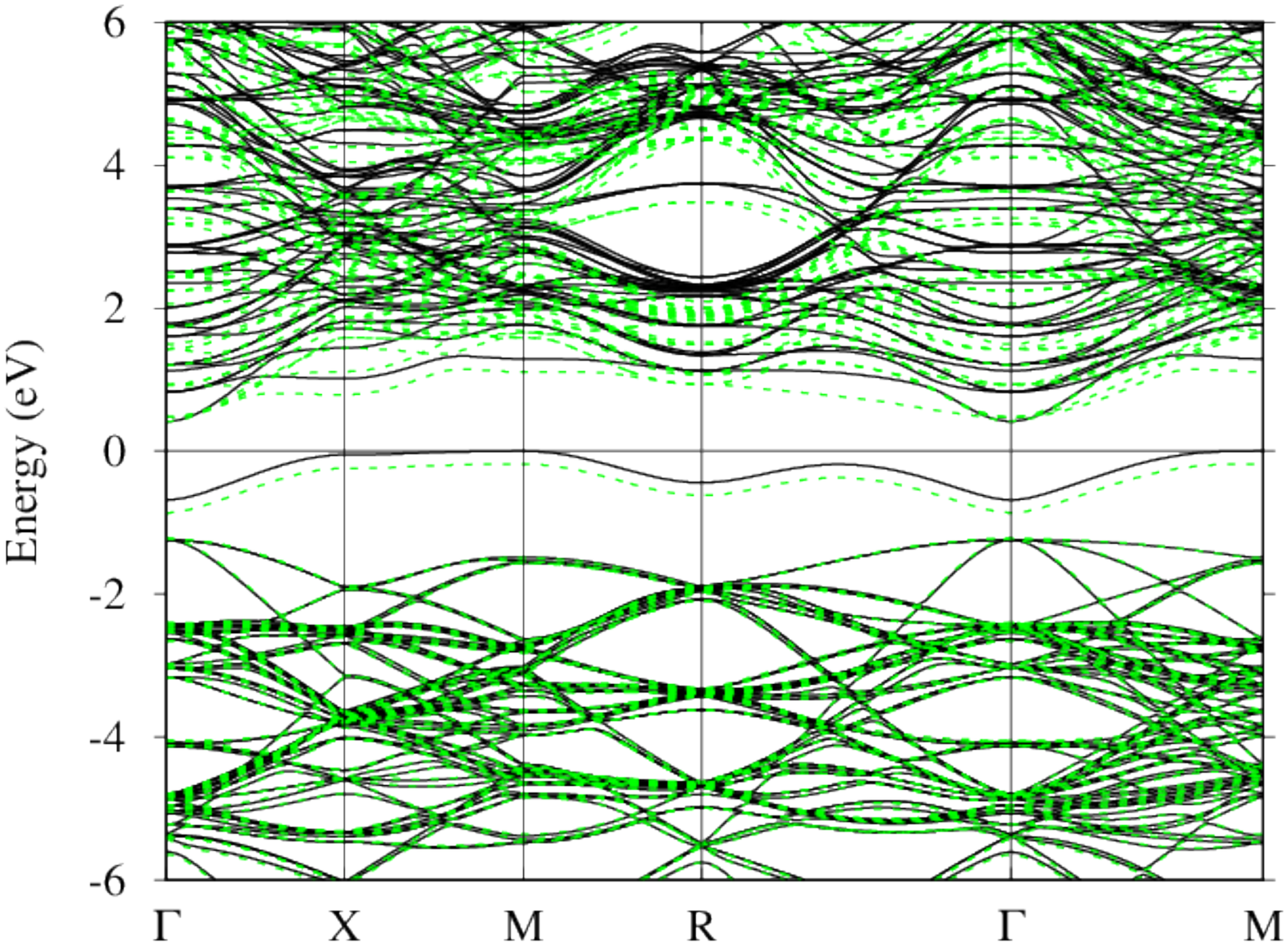}
            \caption{}
            \label{}
        \end{subfigure}
          \hfill
        \begin{subfigure}[b]{0.3\textwidth}
          \centering
            \includegraphics[width=\textwidth]{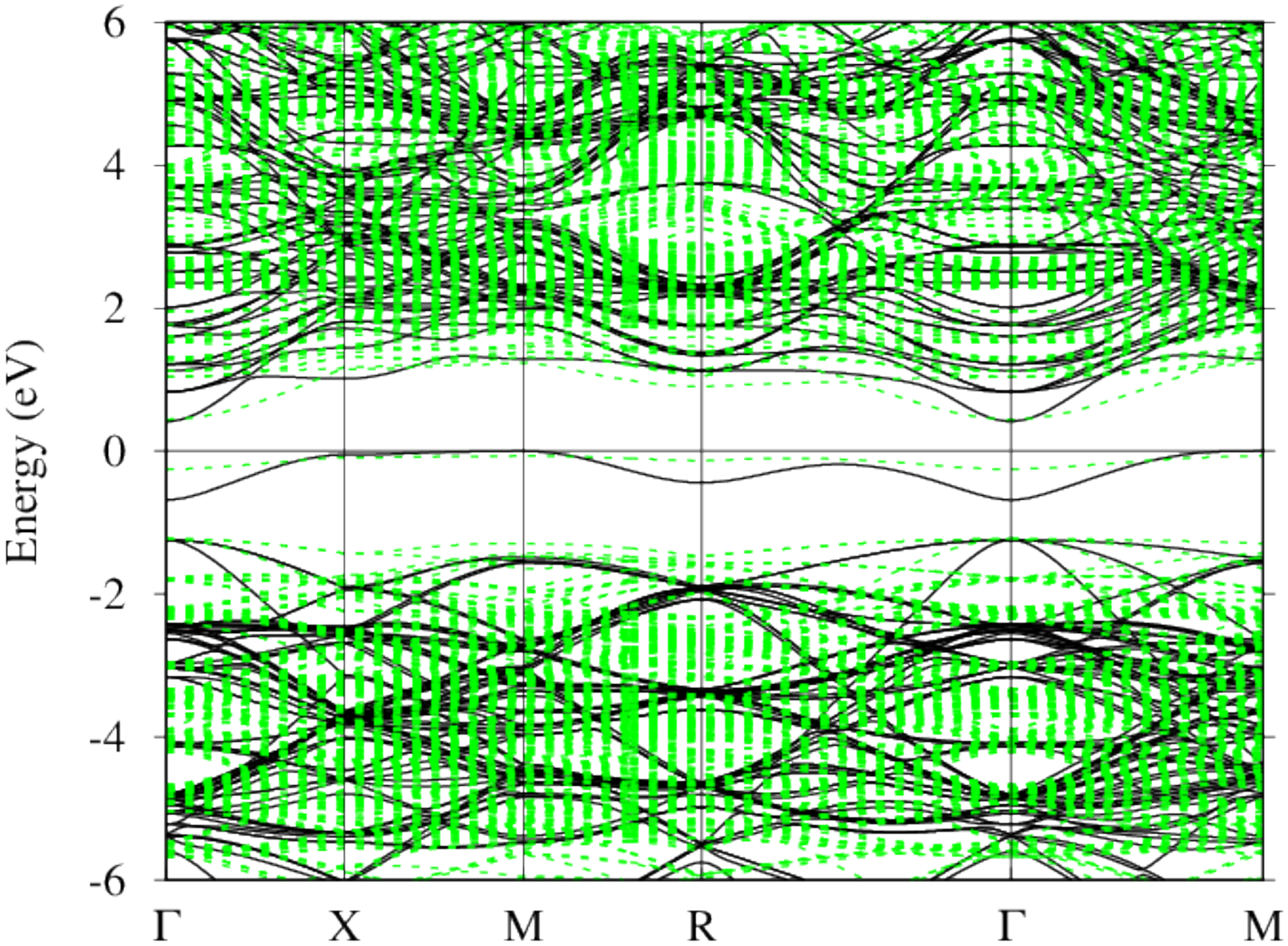}
            \caption{}
            \label{}
        \end{subfigure}
          \caption{``Defect and its nearest neighbors'' defect region in (a) 64$_{8}$, (b)
                   64$_{32}$ and (c) 216$_{8}$ schemes (dashed green line), compared to
                   ``defect only'' defect region in 64$_8$ scheme (solid black line). All
                   structures are in $q=0$ charge state.\label{fig:defnn_def_0}
                   \label{fig:three graphs}}
    \end{figure*}

    \begin{figure*}
      \centering
        \begin{subfigure}[b]{0.3\textwidth}
          \centering
            \includegraphics[width=\textwidth]{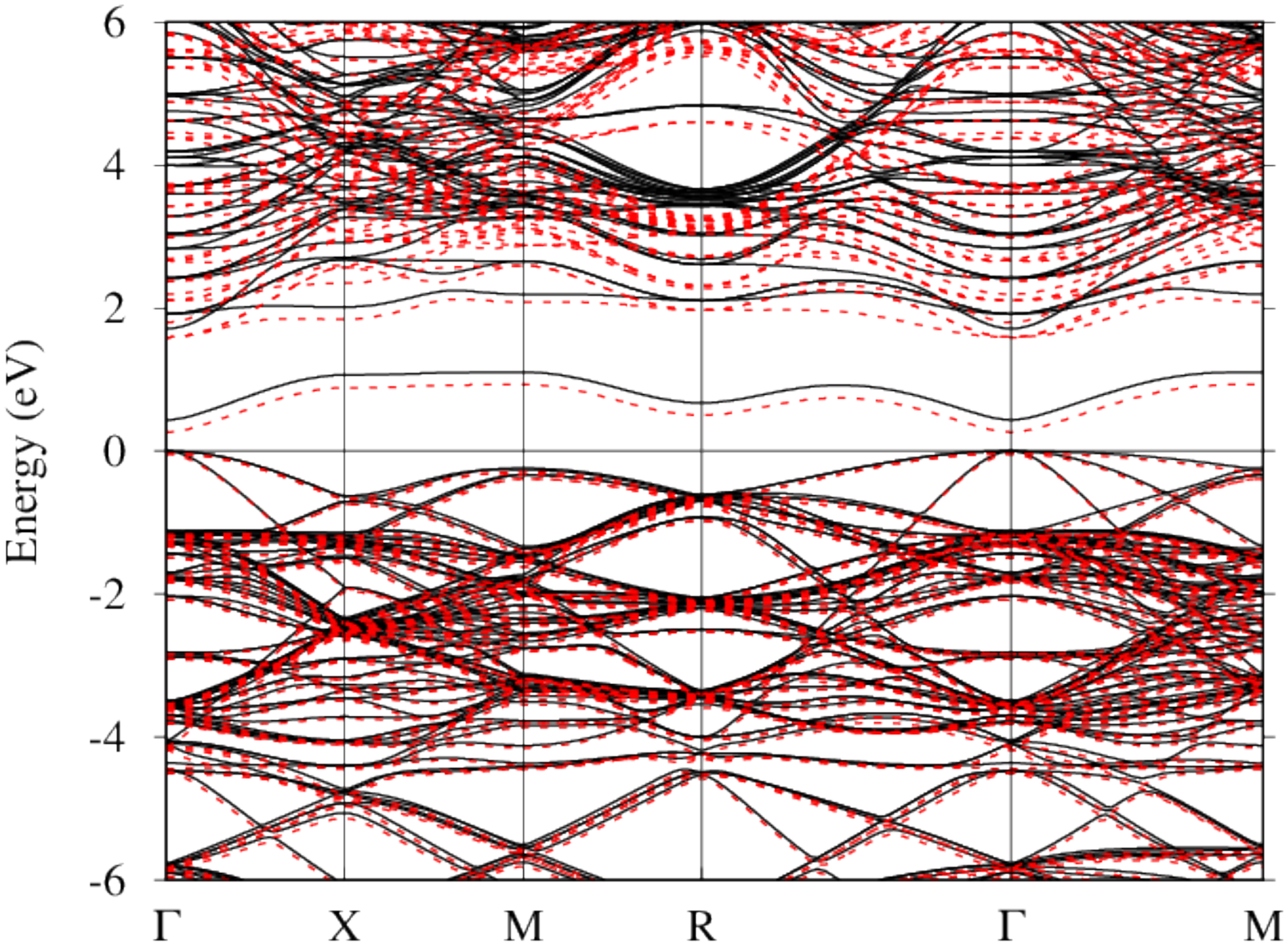}
            \caption{}
            \label{}
        \end{subfigure}
          \hfill
        \begin{subfigure}[b]{0.3\textwidth}
          \centering
            \includegraphics[width=\textwidth]{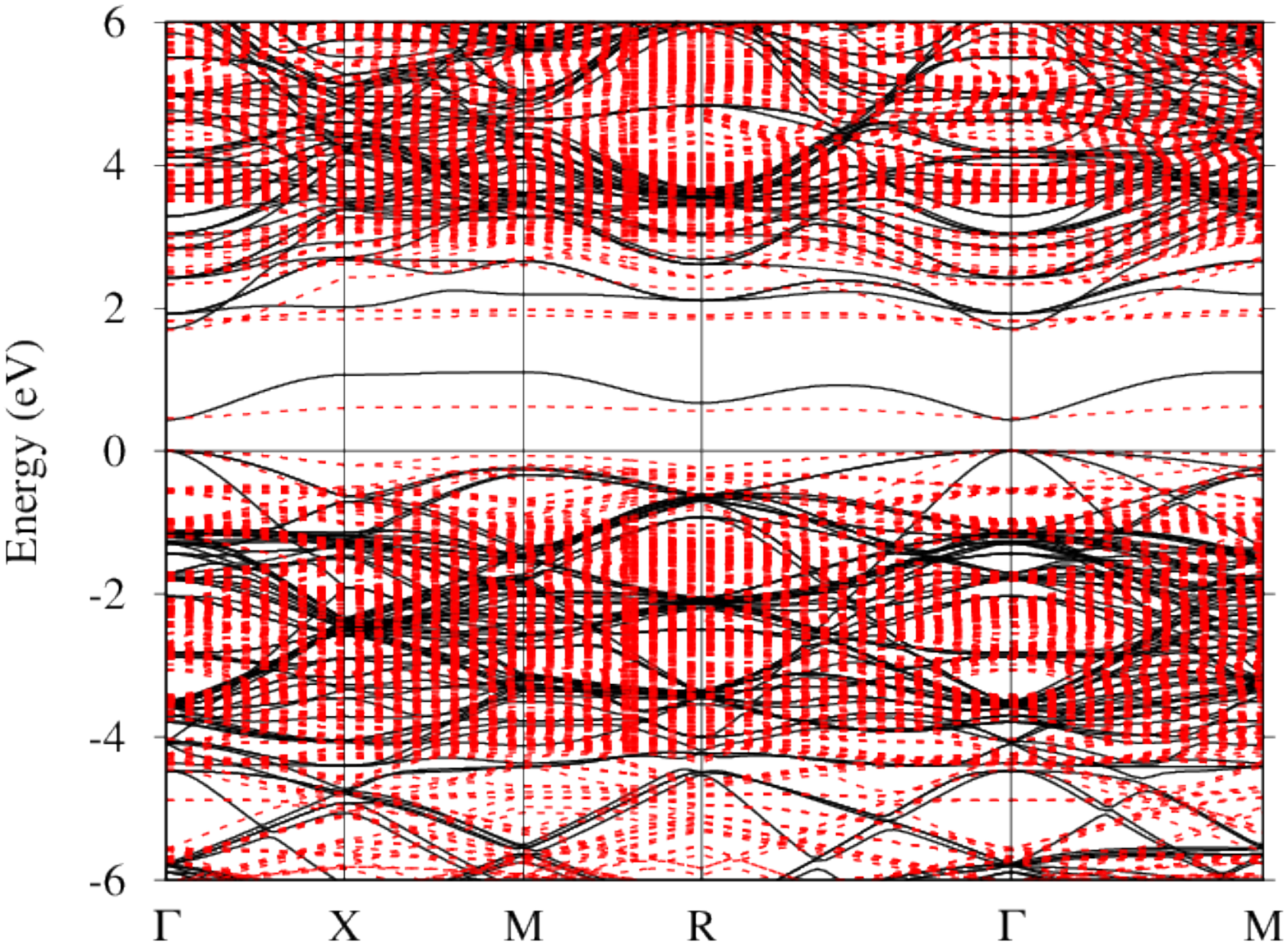}
            \caption{}
            \label{}
        \end{subfigure}
          \caption{Different schemes,namely (a) 64$_{32}$ and (b) 216$_{8}$ schemes
                   (dashed green line), compared to 64$_8$ scheme (solid black line). All
                   structures are in $q=2$ charge state and defect region is only the
                   defect atom.\label{fig:defnn_def_0}\label{fig:def_def_2}}
    \end{figure*}

    \begin{figure*}
      \centering
        \begin{subfigure}[b]{0.3\textwidth}
          \centering
            \includegraphics[width=\textwidth]{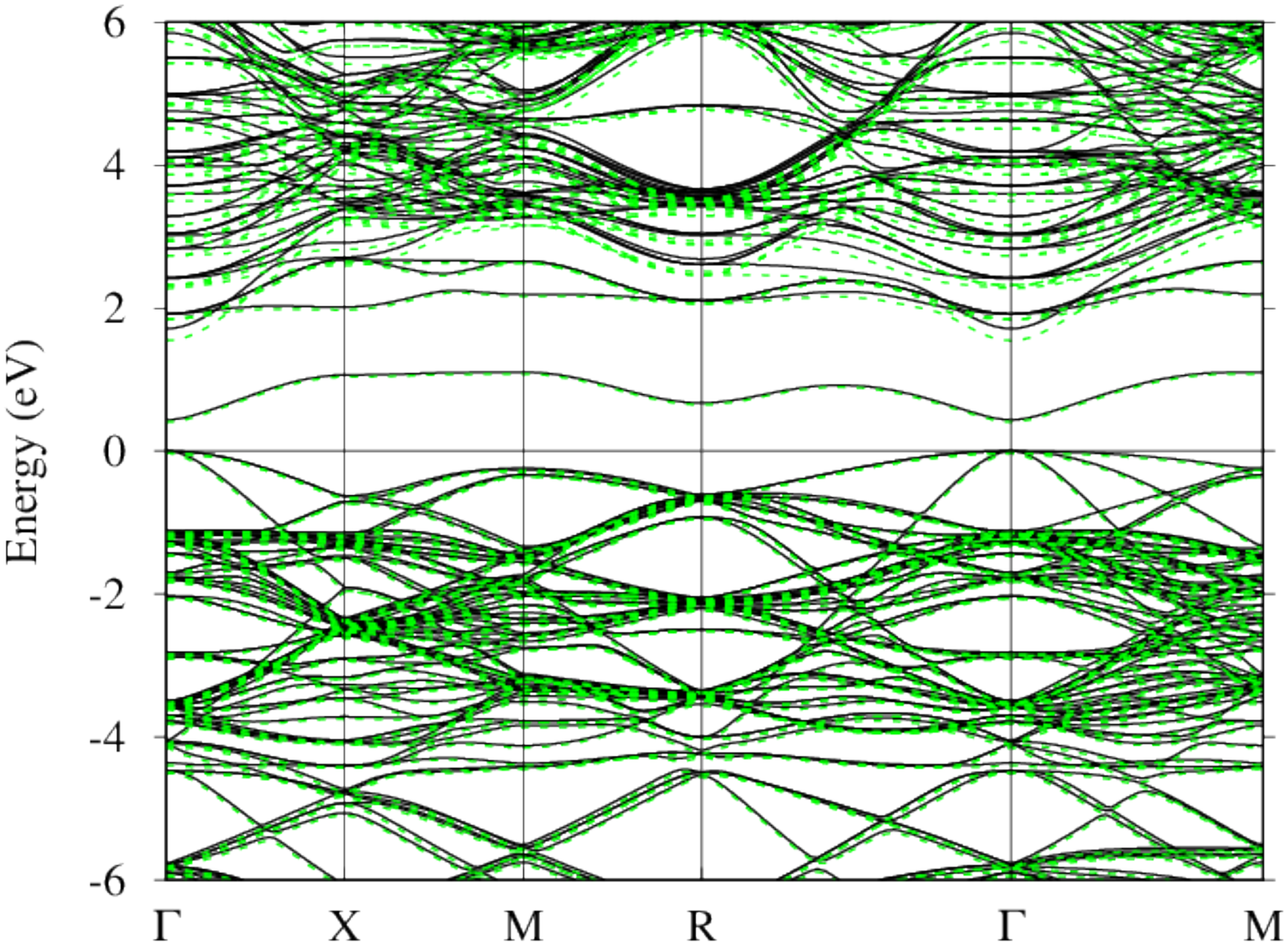}
            \caption{}
            \label{}
        \end{subfigure}
          \hfill
        \begin{subfigure}[b]{0.3\textwidth}
          \centering
            \includegraphics[width=\textwidth]{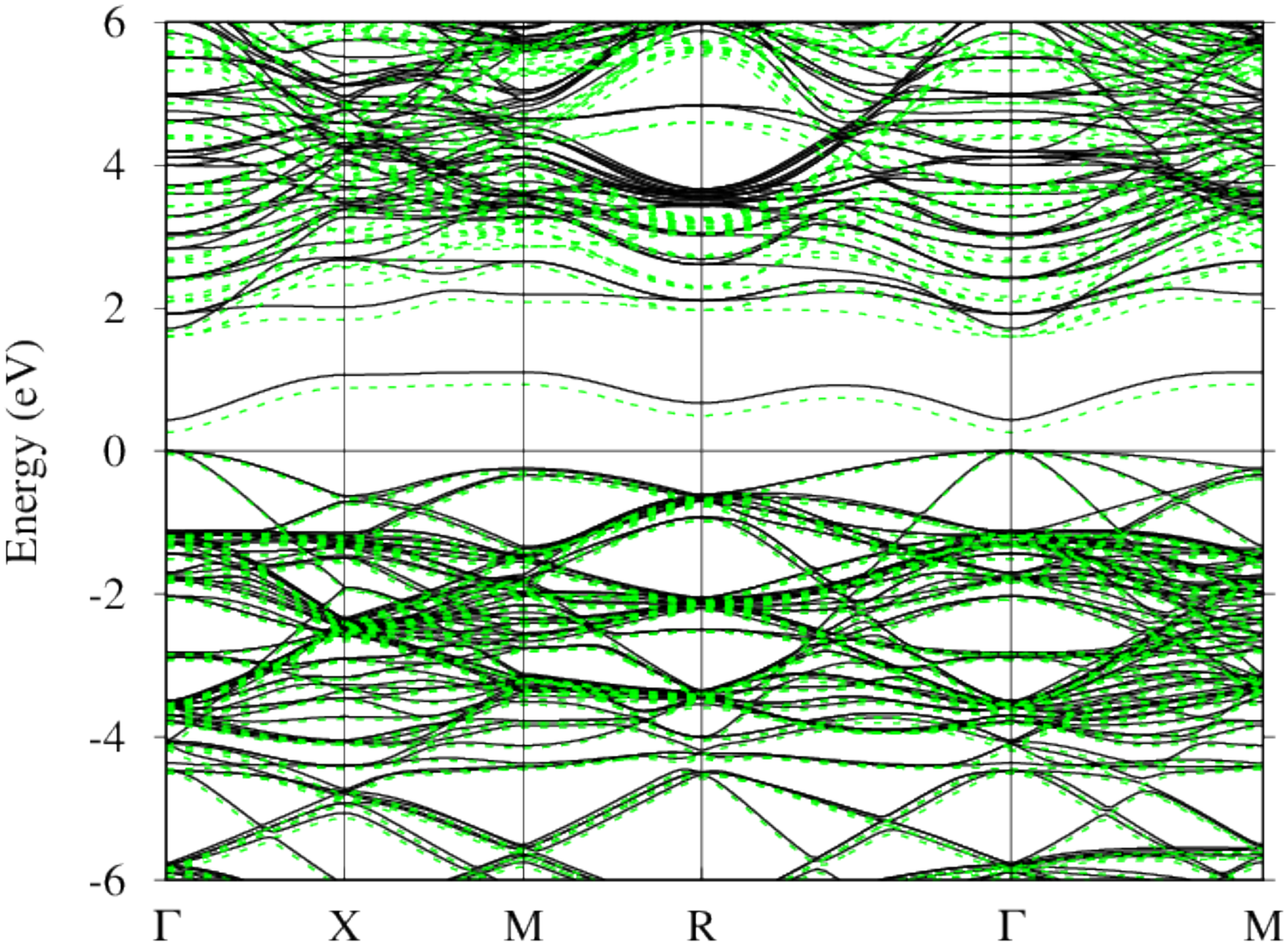}
            \caption{}
            \label{}
        \end{subfigure}
          \hfill
        \begin{subfigure}[b]{0.3\textwidth}
          \centering
            \includegraphics[width=\textwidth]{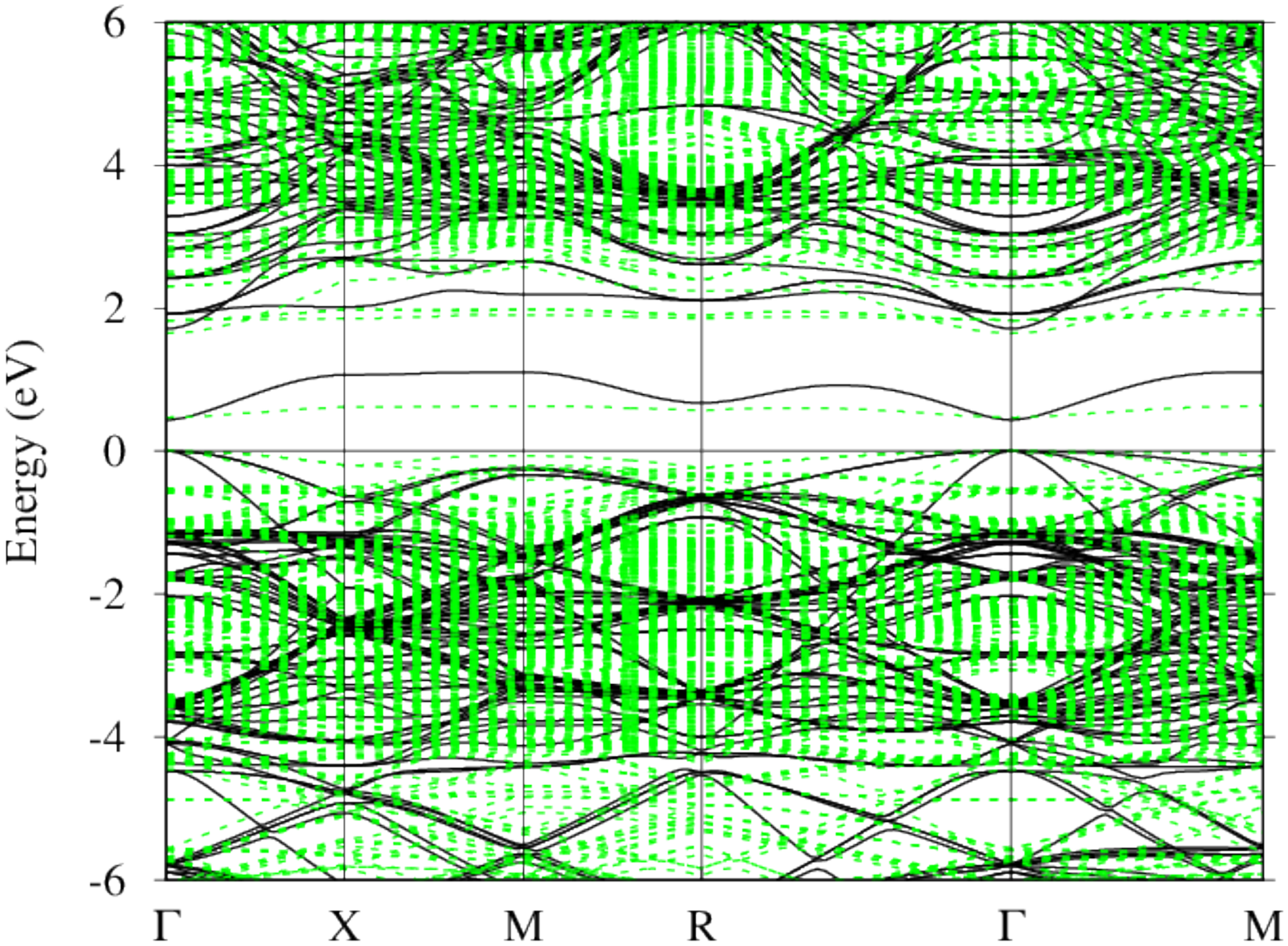}
            \caption{}
            \label{}
        \end{subfigure}
          \caption{``Defect and its nearest neighbors'' defect region in (a) 64$_{8}$, (b)
                   64$_{32}$ and (c) 216$_{8}$ schemes (dashed green line), compared to
                   ``defect only'' defect region in 64$_8$ scheme (solid black line). All
                   structures are in $q=2$ charge state.\label{fig:defnn_def_2}}
    \end{figure*}

    \begin{table}
      \begin{tabular}{lcccccccc}
                             & VBM    & CBM    & t$_2^X$ & a$_1^X$ & t$_2^M$ & a$_1^M$ & a$_1^\Gamma$ & a$_1^{disp}$ \\\hline 
    
        32 atom              & 1.9047 & 4.3332 & 4.6108  & 3.0938  & 4.8652  & 3.1428  & 2.6013       & 0.5415       \\            
        64 atom              & 2.0775 & 4.3305 & 4.7496  & 3.1959  & 5.0026  & 3.2394  & 2.6163       & 0.6231       \\\hline      
        64$^{\dagger}_8$     & 1.8870 & 3.5455 & 4.1441  & 3.0775  & 4.4217  & 3.1305  & 2.4476       & 0.6830       \\            
        64$^{\ast}_8$        & 1.8843 & 3.4054 & 4.1169  & 3.0748  & 4.4162  & 3.1292  & 2.4421       & 0.6871       \\            
        64$^{\dagger}_{32}$  & 1.9020 & 3.5945 & 3.9292  & 2.9115  & 4.2475  & 2.9714  & 2.2952       & 0.6762       \\            
        64$^{\ast}_{32}$     & 1.9047 & 3.5469 & 3.9183  & 2.8870  & 4.2394  & 2.9482  & 2.2653       & 0.6830       \\            
        216$^{\dagger}_{8}$  & 1.9061 & 3.6285 & 4.2883  & 3.0476  & 4.3713  & 3.0680  & 2.8829       & 0.1850       \\            
        216$^{\ast}_{8}$     & 1.9020 & 3.5754 & 4.2734  & 3.0421  & 4.3605  & 3.0639  & 2.8761       & 0.1878       \\\hline       
      \end{tabular}                                    
        \begin{tablenotes}                               
          \small
            \item $^{\dagger}$: Defect region is defect atom itself.
            \item $^{\ast}$: Defect region is defect atom and its nearest neighbors.
        \end{tablenotes}
        \caption{$q=0$ state}
    \end{table}

    \begin{table}
      \begin{tabular}{lcccccccc}
                             & VBM    & CBM    & t$_2^X$ & a$_1^X$ & t$_2^M$ & a$_1^M$ & a$_1^\Gamma$ & a$_1^{disp}$ \\\hline 
    
        32 atom              & 1.9197 & 4.4285 & 4.6747  & 3.4734  & 4.9428  & 3.5183  & 2.9251       & 0.5932       \\             
        64 atom              & 2.0680 & 4.3088 & 4.3537  & 2.7550  & 4.5088  & 2.7836  & 2.1836       & 0.6000       \\\hline       
        64$^{\dagger}_8$     & 1.8843 & 3.5986 & 3.8992  & 2.9482  & 4.0788  & 2.9877  & 2.3210       & 0.6667       \\             
        64$^{\ast}_8$        & 1.8802 & 3.4571 & 3.8843  & 2.9469  & 4.0761  & 2.9877  & 2.3156       & 0.6721       \\             
        64$^{\dagger}_{32}$  & 1.9006 & 3.5265 & 3.7809  & 2.8176  & 4.0190  & 2.8652  & 2.1972       & 0.6680       \\             
        64$^{\ast}_{32}$     & 1.9034 & 3.5414 & 3.7727  & 2.8190  & 4.0217  & 2.8652  & 2.1959       & 0.6694       \\             
        216$^{\dagger}_{8}$  & 1.9211 & 3.6176 & 3.7673  & 2.5278  & 3.8135  & 2.5442  & 2.3836       & 0.1605       \\             
        216$^{\ast}_{8}$     & 1.9156 & 3.5659 & 3.7700  & 2.5333  & 3.8176  & 2.5496  & 2.3877       & 0.1619       \\\hline       
      \end{tabular}                                    
        \begin{tablenotes}                               
          \small
            \item $^{\dagger}$: Defect region is defect atom itself.
            \item $^{\ast}$: Defect region is defect atom and its nearest neighbors.
        \end{tablenotes}
        \caption{$q=2$ state}
    \end{table}
\end{widetext}

\end{document}